\documentclass[12pt,english]{article}
\usepackage{times}
\usepackage[T1]{fontenc}
\usepackage{geometry}
\usepackage{rotating}
\usepackage{graphicx}
\usepackage{setspace}
\usepackage{amssymb}

\makeatletter

\providecommand{\tabularnewline}{\\}

\usepackage{bm}

\usepackage{babel}
\makeatother
\begin{document}

\section*{\noindent Holographic Smart \emph{EM} Skins for Advanced Beam Power
Shaping in Next Generation Wireless Environments}

\noindent \vfill

\noindent G. Oliveri,$^{(1)}$ \emph{Senior Member, IEEE}, P. Rocca,
$^{(1)(2)}$ \emph{Senior Member, IEEE}, M. Salucci,$^{(1)}$ \emph{Member,
IEEE}, and A. Massa,$^{(1)(3)(4)}$ \emph{Fellow, IEEE}

\noindent \vfill

\noindent {\footnotesize ~}{\footnotesize \par}

\noindent {\footnotesize $^{(1)}$} \emph{\footnotesize CNIT - \char`\"{}University
of Trento\char`\"{} Research Unit }{\footnotesize \par}

\noindent {\footnotesize Via Sommarive 9, 38123 Trento - Italy}{\footnotesize \par}

\noindent \textit{\emph{\footnotesize E-mail:}} {\footnotesize \{}\emph{\footnotesize giacomo.oliveri,
paolo.rocca, marco.salucci, andrea.massa}{\footnotesize \}@}\emph{\footnotesize unitn.it}{\footnotesize \par}

\noindent {\footnotesize Website:} \emph{\footnotesize www.eledia.org/eledia-unitn}{\footnotesize \par}

\noindent {\footnotesize ~}{\footnotesize \par}

\noindent {\footnotesize $^{(2)}$} \emph{\footnotesize ELEDIA Research
Center} {\footnotesize (}\emph{\footnotesize ELEDIA}{\footnotesize @}\emph{\footnotesize XIDIAN}
{\footnotesize - Xidian University)}{\footnotesize \par}

\noindent {\footnotesize P.O. Box 191, No.2 South Tabai Road, 710071
Xi'an, Shaanxi Province - China }{\footnotesize \par}

\noindent {\footnotesize E-mail:} \emph{\footnotesize paolo.rocca@xidian.edu.cn}{\footnotesize \par}

\noindent {\footnotesize Website:} \emph{\footnotesize www.eledia.org/eledia-xidian}{\footnotesize \par}

\noindent {\footnotesize ~}{\footnotesize \par}

\noindent {\footnotesize $^{(3)}$} \emph{\footnotesize ELEDIA Research
Center} {\footnotesize (}\emph{\footnotesize ELEDIA}{\footnotesize @}\emph{\footnotesize UESTC}
{\footnotesize - UESTC)}{\footnotesize \par}

\noindent {\footnotesize School of Electronic Engineering, Chengdu
611731 - China}{\footnotesize \par}

\noindent \textit{\emph{\footnotesize E-mail:}} \emph{\footnotesize andrea.massa@uestc.edu.cn}{\footnotesize \par}

\noindent {\footnotesize Website:} \emph{\footnotesize www.eledia.org/eledia}{\footnotesize -}\emph{\footnotesize uestc}{\footnotesize \par}

\noindent {\footnotesize ~}{\footnotesize \par}

\noindent {\footnotesize $^{(4)}$} \emph{\footnotesize ELEDIA Research
Center} {\footnotesize (}\emph{\footnotesize ELEDIA@TSINGHUA} {\footnotesize -
Tsinghua University)}{\footnotesize \par}

\noindent {\footnotesize 30 Shuangqing Rd, 100084 Haidian, Beijing
- China}{\footnotesize \par}

\noindent {\footnotesize E-mail:} \emph{\footnotesize andrea.massa@tsinghua.edu.cn}{\footnotesize \par}

\noindent {\footnotesize Website:} \emph{\footnotesize www.eledia.org/eledia-tsinghua}{\footnotesize \par}

\noindent \vfill

\newpage
\section*{Holographic Smart \emph{EM} Skins for Advanced Beam Power Shaping
in Next Generation Wireless Environments}

~

~

~

\begin{flushleft}G. Oliveri, P. Rocca, M. Salucci, and A. Massa\end{flushleft}

\vfill

\begin{abstract}
\noindent An innovative approach for the synthesis of inexpensive
holographic smart electromagnetic (\emph{EM}) skins with advanced
beamforming features is proposed. The complex multi-scale smart skin
design is formulated within the Generalized Sheet Transition Condition
(\emph{GSTC}) framework as a combination of a mask-constrained isophoric
inverse source problem and a micro-scale susceptibility dyadic optimization.
The solution strategy integrates a local search procedure based on
the iterative projection technique (\emph{IPT}) and a \emph{System-by-Design}
(\emph{SbD})-based optimization loop for the identification of optimal
metasurface descriptors matching the desired surface currents. The
performance and the efficiency of the proposed approach are assessed
in a set of representative test cases concerned with different smart
skin apertures and target pattern masks.

\vfill
\end{abstract}
\noindent \textbf{Key words}: Smart Skins; \emph{EM} Holography; Next-Generation
Communications; Iterative Projection Method; System-by-Design; Metasurfaces;
Metamaterials.

\newpage
\section{Introduction and Rationale\label{sec:Introduction}}

\noindent The next generation of wireless cellular systems is envisaged
to fulfil unprecedented requirements in terms of data transfer speed,
flexibility, coverage, reliability, and quality of service \cite{Liaskos 2018}-\cite{Di Renzo 2020b}.
The need to meet such ambitious expectations, while still relying
on cost-effective and efficient technologies, is motivating a deep
re-visitation of the paradigms currently adopted in the design and
the deployment of wireless communication systems \cite{Basar 2019}-\cite{Di Renzo 2020b}.
As a matter of fact, the transition between subsequent wireless communication
generations has traditionally consisted in upgrading the technological
solutions of the user terminals as well as of the provider base stations
and network \cite{Basar 2019}-\cite{Di Renzo 2020b}. On the contrary,
the propagation environment has been considered as a fundamental,
but essentially uncontrollable, element/actor of the wireless scenario
\cite{Basar 2019}-\cite{Di Renzo 2020b}. This viewpoint is being
completely overrun by the emerging paradigm of the Smart Electromagnetic
Environment (\emph{SEE}) \cite{Liaskos 2018}-\cite{Massa 2021}.
The transformative \emph{SEE} vision originates from the key idea
that the wireless propagation can be partially controlled by properly
{}``tailoring'' the reflection by buildings and urban structures
\cite{Basar 2019}-\cite{Di Renzo 2020b}\cite{Xu 2018}\cite{Massa 2021}.
In the \emph{SEE} scenario, the environment is no longer an uncontrollable
part of a wireless system, but rather it can cooperatively support
the propagation to improve the coverage, the data rate, and the network
reliability without the need to install additional base stations \cite{Di Renzo 2019}-\cite{Di Renzo 2020b}\cite{Massa 2021}\cite{Huang 2019}.

\noindent The revolutionary potentialities of the \emph{SEE} are based
on the exploitation of thin metasurfaces operating as smart electromagnetic
(\emph{EM}) \emph{}skins \cite{Di Renzo 2019}\cite{Tang 2019}\cite{Massa 2021}\cite{Pitilakis 2021}\cite{Diaz 2021}.
In short, such a technology enables the meta-atomic manipulation of
the reflected/transmitted wavefronts to overcome the traditional Snell's
laws \cite{Di Renzo 2019}\cite{Pitilakis 2021}-\cite{Yang 2019}.
This implies a wide set of unconventional phenomena such as anomalous
reflections, focusing/lensing effects, polarization control, perfect
absorption, holography, non-reciprocity, extreme energy accumulation,
and enhanced security \cite{Di Renzo 2019}\cite{Pitilakis 2021}-\cite{Yang 2019}.
Depending on manipulation properties and technological constraints,
different classes of smart skins have been considered. On the one
hand, dynamically adjustable artificial materials operating as \emph{Reconfigurable
Intelligent Surfaces} (\emph{RIS}) give the control of the reflected
wave properties in real time \cite{Basar 2019}-\cite{Di Renzo 2020b}\cite{Pitilakis 2021},
but at the cost of non-negligible implementation complexity, costs,
and power consumption. On the other hand, static passive smart \emph{EM}
skins (\emph{SPSS}s) virtually imply no running costs after installation
and they potentially have advanced beamforming capabilities \cite{Massa 2021}\cite{Diaz 2021}-\cite{Oliveri 2015b}.
However, the design of \emph{SPSS}s is very challenging because of
the reduced set of degrees-of-freedom (\emph{DoF}s) more severely
constrained in terms of final layout complexity (e.g., passive instead
of active, static instead of reconfigurable). As a matter of fact,
let us notice that the arising wave manipulation device must be simple,
light, and inexpensive to manufacture despite its wide \emph{EM} size
comprising hundreds of thousands of unit cells. Moreover, it is required
to have a careful and robust macro-scale beam control for enabling
beam focusing since no adjustment (e.g., no calibration or real-time
control) to the reflection properties is possible after the prototyping.
Furthermore, the \emph{SPSS}s must be large enough to guarantee that
an adequate level of power is reflected towards the whole coverage
area since no {}``per-user'' beam is allowed. This results in a
huge number of micro-scale design descriptors to be optimized during
the \emph{SPSS} synthesis. Finally, unlike reflectarray engineering,
the \emph{SPSS}s performance must be yielded with a limited control
on the surface orientation with respect to the incident direction
of the illuminating beam.

\noindent The objective of this paper is to give some indications
on the feasibility of simple and inexpensive holographic \emph{SPSS}s
suitable for advanced wave manipulations and beamforming. Towards
this end, the complex multi-scale \emph{EM} design problem at hand
is firstly formulated within the \emph{Generalized Sheet Transition
Condition} (\emph{GSTC}) theoretical framework \cite{Yang 2019}\cite{Ricoy 1990}-\cite{Achouri 2015}.
Then, a phase-only inverse source (\emph{IS}) approach is adopted
to generalize the concepts introduced in \cite{Salucci 2018} for
reflectarray engineering to the synthesis of a holographic metasurface
working in the \emph{SEE} scenario. The footprint coverage capabilities
of the smart \emph{EM} skin are successively optimized by combining
(\emph{a}) a local search approach, based on the \emph{Iterative Projection
Technique} (\emph{IPT}) \cite{Rocca 2009b} and (\emph{b}) a customized
version of the \emph{System-by-Design} (\emph{SbD}) paradigm. More
in detail, the former (\emph{a}) is aimed at deducing the reference/ideal
surface currents affording the user-defined footprint pattern, while
the other (\emph{b}) is devoted to set the descriptors (i.e., the
\emph{DoF}s) of the \emph{SPSS} for matching those reference currents.
Such methodological choices, to implement a synthesis method for \emph{SPSS}s,
are driven by (\emph{i}) the accuracy of the \emph{GSTC} theory in
accounting for the complex \emph{EM} response of smart skins in the
\emph{SEE} framework \cite{Di Renzo 2020}\cite{Di Renzo 2020b}\cite{Yang 2019},
(\emph{ii}) the effectiveness of the \emph{SbD} in handling complex
multi-scale design problems \cite{Massa 2014}-\cite{Oliveri 2021},
and (\emph{iii}) the intrinsic advantages of exploiting an \emph{IS}
formulation when determining surface currents \cite{Salucci 2018}
(e.g., the possibility to introduce non-radiating components for fitting
further user-requirements in terms of manufacturing, as well). Consequently,
the main innovative contributions of this work lie in (\emph{i}) the
customization of the \emph{SbD} paradigm within the \emph{GSTC} framework,
(\emph{ii}) the combination of the \emph{SbD}-based technique and
of an \emph{IPT}-based source synthesis process to afford complex
pattern footprints with simple and inexpensive \emph{SPSS} layouts,
and (\emph{iii}) the numerical assessment of the effectiveness of
the proposed approach as well as of the feasibility of holographic
\emph{SPSS}s able to generate complex footprints.

\noindent The outline of the paper is as follows. The problem of designing
a holographic smart passive \emph{EM} skin fitting user-defined requirements
is formulated in Sect. \ref{sec:Problem-Formulation}. Section \ref{sec:Method}
details the proposed synthesis approach. Selected numerical results,
drawn from an extensive numerical validation, are illustrated in Sect.
\ref{sec:Numerical-Analysis-and}. Finally, some concluding remarks
are reported (Sect. \ref{sec:Conclusions-and-Remarks}).

\section{\noindent Problem Formulation\label{sec:Problem-Formulation} }

\noindent With reference to the scenario in Fig. 1 and without loss
of generality, let us consider a \emph{SPSS} composed by $P\times Q$
meta-film unit cells located at the positions \{$\mathbf{r}_{pq}\in\Omega$;
$p=1,...,P$; $q=1,...,Q$\}, $\Omega$ being the smart skin aperture/support,
and illuminated by an incident plane wave impinging from the angular
direction $\left(\theta^{inc},\varphi^{inc}\right)$ whose associated
electric and magnetic fields are \cite{Lindell 2019}\cite{Osipov 2017}\begin{equation}
\mathbf{E}^{inc}\left(\mathbf{r}\right)\triangleq\left(E_{\bot}^{inc}\widehat{\mathbf{e}}_{\bot}+E_{\parallel}^{inc}\widehat{\mathbf{e}}_{\parallel}\right)\exp\left(-j\mathbf{k}^{inc}\cdot\mathbf{r}\right)\label{eq:incident wave}\end{equation}
and $\mathbf{H}^{inc}\left(\mathbf{r}\right)\triangleq\frac{1}{\eta_{0}k_{0}}\mathbf{k}^{inc}\times\mathbf{E}^{inc}\left(\mathbf{r}\right)$,
respectively, $\mathbf{k}^{inc}$ being the incident wave vector\begin{equation}
\mathbf{k}^{inc}\triangleq-k_{0}\left[\sin\left(\theta^{inc}\right)\cos\left(\varphi^{inc}\right)\widehat{\mathbf{x}}+\sin\left(\theta^{inc}\right)\sin\left(\varphi^{inc}\right)\widehat{\mathbf{y}}+\cos\left(\theta^{inc}\right)\widehat{\mathbf{z}}\right],\label{eq:incident wave vector}\end{equation}
 while $\mathbf{r}=\left(x,y,z\right)$ is the metasurface local coordinate,
$k_{0}$ and $\eta_{0}$ being the free-space wavenumber and intrinsic
impedance, respectively. Moreover, $\widehat{\mathbf{e}}_{\bot}=\frac{\mathbf{k}^{inc}\times\widehat{\mathbf{n}}}{\left|\mathbf{k}^{inc}\times\widehat{\mathbf{n}}\right|}$
and $\widehat{\mathbf{e}}_{\parallel}=\frac{\widehat{\mathbf{e}}_{\bot}\times\mathbf{k}^{inc}}{\left|\widehat{\mathbf{e}}_{\bot}\times\mathbf{k}^{inc}\right|}$
are the {}``perpendicular'' and {}``parallel'' unit vectors (i.e.,
\emph{TE} and \emph{TM} mode), respectively, while $E_{\bot}^{inc}$
and $E_{\parallel}^{inc}$ are the corresponding complex-valued coefficients,
$\widehat{\mathbf{n}}$ is the normal to the smart skin surface, and
$\left|\cdot\right|$ is the vector magnitude operator. In far-field,
the electric field reflected by the \emph{SPSS} is given by \cite{Osipov 2017}

\noindent \begin{equation}
\mathbf{E}^{FF}\left(\mathbf{r}\right)\approx\frac{jk_{0}}{4\pi}\frac{\exp\left(-jk_{0}\left|\mathbf{r}\right|\right)}{\left|\mathbf{r}\right|}\int_{\Omega}\left\{ \widehat{\mathbf{r}}\times\left[\eta_{0}\widehat{\mathbf{r}}\times\mathbf{J}^{e}\left(\widetilde{\mathbf{r}}\right)+\mathbf{J}^{m}\left(\widetilde{\mathbf{r}}\right)\right]\exp\left(jk_{0}\widehat{\mathbf{r}}\cdot\widetilde{\mathbf{r}}\right)\right\} \mathrm{d}\widetilde{\mathbf{r}}\label{eq:far field}\end{equation}
where $\widehat{\mathbf{r}}=\frac{\mathbf{r}}{\left|\mathbf{r}\right|}$.
Moreover, the effective equivalent electric/magnetic surface current
\cite{Yang 2019}\cite{Osipov 2017}, $\mathbf{J}^{e}\left(\mathbf{r}\right)$/$\mathbf{J}^{m}\left(\mathbf{r}\right)$,
is computed according to the \emph{GSTC} as follows \cite{Yang 2019}\cite{Achouri 2015}:\begin{equation}
\mathbf{J}^{e}\left(\mathbf{r}\right)=j\omega\mathbf{B}_{t}^{e}\left(\mathbf{r}\right)-\widehat{\mathbf{n}}\times\nabla_{t}B_{n}^{m}\left(\mathbf{r}\right)\quad\mathbf{r}\in\Omega\label{eq:GSTC Je}\end{equation}
\begin{equation}
\mathbf{J}^{m}\left(\mathbf{r}\right)=j\omega\mu_{0}\mathbf{B}_{t}^{m}\left(\mathbf{r}\right)+\widehat{\mathbf{n}}\times\nabla_{t}\frac{B_{n}^{e}\left(\mathbf{r}\right)}{\varepsilon_{0}}\quad\mathbf{r}\in\Omega\label{eq:GSTC Jm}\end{equation}
where $\varepsilon_{0}$ and $\mu_{0}$ are the free-space permittivity
and permeability, respectively, while $\mathbf{B}^{e}\left(\mathbf{r}\right)=\mathbf{B}_{t}^{e}\left(\mathbf{r}\right)+B_{n}^{e}\left(\mathbf{r}\right)\widehat{\mathbf{n}}$
and $\mathbf{B}^{m}\left(\mathbf{r}\right)=\mathbf{B}_{t}^{m}\left(\mathbf{r}\right)+B_{n}^{m}\left(\mathbf{r}\right)\widehat{\mathbf{n}}$
are the electric and the magnetic polarization surface densities whose
expressions, under the local periodicity assumption and considering
(sufficiently) symmetric unit cells, are \cite{Yang 2019}\cite{Achouri 2015}\begin{equation}
\mathbf{B}^{e}\left(\mathbf{r}\right)\approx\sum_{p=1}^{P}\sum_{q=1}^{Q}\left[\varepsilon_{0}\overline{\overline{\chi}}\left(\mathbf{d}_{pq}\right)\cdot\mathbf{E}_{pq}^{ave}\right]\Pi^{pq}\left(\mathbf{r}\right)\quad\mathbf{r}\in\Omega\label{eq:electric polariz}\end{equation}
\begin{equation}
\mathbf{B}^{m}\left(\mathbf{r}\right)\approx\sum_{p=1}^{P}\sum_{q=1}^{Q}\left[\overline{\overline{\xi}}\left(\mathbf{d}_{pq}\right)\cdot\mathbf{H}_{pq}^{ave}\right]\Pi^{pq}\left(\mathbf{r}\right)\quad\mathbf{r}\in\Omega.\label{eq:magnetic polar}\end{equation}
where $\overline{\overline{\chi}}\left(\mathbf{d}_{pq}\right)\triangleq\sum_{i=x,y,z}\chi_{ii}\left(\mathbf{d}_{pq}\right)\widehat{i}\widehat{i}$
and $\overline{\overline{\xi}}\left(\mathbf{d}_{pq}\right)\triangleq\sum_{i=x,y,z}\xi_{ii}\left(\mathbf{d}_{pq}\right)\widehat{i}\widehat{i}$
are the diagonal tensors of the electric and the magnetic local surface
susceptibilities of the ($p$, $q$)-th ($p=1,...,P$; $q=1,...,Q$)
unit cell described by the $L$-size set $\mathbf{d}_{pq}\triangleq\left\{ d_{pq}^{\left(l\right)},l=1,...,L\right\} $,
while $\Pi^{pq}\left(\mathbf{r}\right)\triangleq\left\{ 1\, if\,\mathbf{r}\in\Omega_{pq},\,0\, if\,\mathbf{r}\notin\Omega_{pq}\right\} $
is the basis function defined on the ($p$, $q$)-th ($p=1,...,P$;
$q=1,...,Q$) cell support $\Omega_{pq}$ ($\sum_{p=1}^{P}\sum_{q=1}^{Q}\Omega_{pq}=\Omega$).
Moreover, $\bm{\Psi}_{pq}^{ave}$ ($\bm{\Psi}=\left\{ \mathbf{E},\mathbf{H}\right\} $)
is the surface averaged field defined as \cite{Yang 2019}\begin{equation}
\bm{\Psi}_{pq}^{ave}\triangleq\frac{\int_{\Omega_{pq}}\left[\bm{\Psi}^{inc}\left(\mathbf{r}\right)+\bm{\Psi}^{ref}\left(\mathbf{r}\right)\right]d\mathbf{r}}{2\times\int_{\Omega_{pq}}d\mathbf{r}},\label{eq:field average}\end{equation}
where the local reflected electric/magnetic field $\bm{\Psi}^{ref}$
is given by

\noindent \begin{equation}
\bm{\Psi}^{ref}\left(\mathbf{r}\right)=\overline{\overline{\Gamma}}\left[\overline{\overline{\chi}}\left(\mathbf{d}_{pq}\right),\overline{\overline{\xi}}\left(\mathbf{d}_{pq}\right)\right]\cdot\bm{\Psi}^{inc}\left(\mathbf{r}\right)\label{eq:reflected field}\end{equation}
where $\overline{\overline{\Gamma}}$ is the local reflection tensor
\cite{Yang 2019}\begin{equation}
\overline{\overline{\Gamma}}\left[\overline{\overline{\chi}},\overline{\overline{\xi}}\right]\triangleq\left[\begin{array}{cc}
\Gamma_{\bot\bot} & \Gamma_{\parallel\bot}\\
\Gamma_{\bot\parallel} & \Gamma_{\parallel\parallel}\end{array}\right].\label{eq:Gamma Local}\end{equation}
According to the above derivation, the design of the holographic \emph{SPSS}
able to generate a desired footprint mask in a \emph{Coverage Region}
$\Xi$ can be carried out by solving the following two sub-problems:

\begin{quotation}
\noindent \textbf{\emph{\underbar{Sub-Problem 1}}} - The synthesis
of the \emph{ideal}/\emph{reference} surface currents, \{$\left[\mathbf{J}^{w}\left(\mathbf{r}\right)\right]^{*}$;
$w=\left\{ e,m\right\} $\}, that radiate a far-field pattern (\ref{eq:far field})
fitting in $\Xi$ (i.e., $\mathbf{r}\in\Xi$) the pattern requirements
expressed in terms of lower, $\mathcal{L}\left(\mathbf{r}\right)$,
and upper, $\mathcal{U}\left(\mathbf{r}\right)$, user-defined footprint
power masks\begin{equation}
\mathcal{L}\left(\mathbf{r}\right)\leq\left|\left[\mathbf{E}^{FF}\left(\mathbf{r}\right)\right]^{*}\right|^{2}\leq\mathcal{U}\left(\mathbf{r}\right);\label{eq:mask objective}\end{equation}
\textbf{\emph{\underbar{Sub-Problem 2}}} - The retrieval of the optimal
setup of the \emph{SPSS} descriptors, $\mathcal{D}^{opt}=\left\{ \mathbf{d}_{pq}^{opt};\, p=1,...,P;\, q=1,...,Q\right\} $
so that the \emph{target} surface currents computed by substituting
(\ref{eq:electric polariz}) and (\ref{eq:magnetic polar}) in (\ref{eq:GSTC Je})
and (\ref{eq:GSTC Jm}) are as close as possible to the ideal ones,
\{ $\left[\mathbf{J}^{w}\left(\mathbf{r}\right)\right]^{*}$; $w=\left\{ e,m\right\} $\},
derived in the \emph{Sub-Problem 1}\begin{equation}
\mathcal{D}^{opt}=\arg\left\{ \min_{\mathcal{D}}\left[\upsilon\left(\mathbf{J}^{w}\left(\mathbf{r}\right);\,\left[\mathbf{J}^{w}\left(\mathbf{r}\right)\right]^{*}\right)\right]\right\} \label{eq:current matching}\end{equation}
where $\upsilon\left(\mathbf{J}^{w}\left(\mathbf{r}\right);\,\left[\mathbf{J}^{w}\left(\mathbf{r}\right)\right]^{*}\right)\triangleq\frac{\sum_{w=\left\{ e,m\right\} }\left\Vert \left[\mathbf{J}^{w}\left(\mathbf{r}\right)\right]^{*}-\mathbf{J}^{w}\left(\mathbf{r}\right)\right\Vert }{\sum_{w=\left\{ e,m\right\} }\left\Vert \left[\mathbf{J}^{w}\left(\mathbf{r}\right)\right]^{*}\right\Vert }$
is the surface currents fidelity index, while $\mathcal{D}\triangleq\left\{ \mathbf{d}_{pq};p=1,...,P,\, q=1,...,Q\right\} $
and $\left\Vert \cdot\right\Vert $ stands for the $\ell_{2}$-norm
operator.
\end{quotation}
\noindent It is worth to point out the multi-scale nature of the overall
\emph{SPSS} synthesis, which is aimed at fulfilling macro-scale objectives
{[}i.e., footprint pattern features according to (\ref{eq:mask objective}){]},
while acting at the unit-cell level by optimizing the small-scale
descriptors of the \emph{SPSS} unit cells, \{$d_{pq}^{\left(l\right)}$;
$l=1,...,L$; $p=1,...,P$; $q=1,...,Q$\}. Moreover, it is very important
to take into account, when defining the synthesis strategy, that the
computational complexity of the problem at hand is very high since
the total number of descriptors, $N_{\mathcal{D}}$ ($N_{\mathcal{D}}\triangleq P\times Q\times L$),
quickly grows with the smart skin aperture and the complexity of the
shape of the \emph{SPSS} unit cell.

\section{\noindent Synthesis Procedure \label{sec:Method}}

\noindent To solve the synthesis problem formulated in Sect. \ref{sec:Problem-Formulation}
in terms of two sub-problems, a combination of \emph{ad-hoc} customized
techniques is considered and detailed in the following. As for the
\emph{IS} concerned with the synthesis of the \emph{ideal} surface
currents, \{$\left[\mathbf{J}^{w}\left(\mathbf{r}\right)\right]^{*}$,
$w=\left\{ e,m\right\} $\}, according to (\ref{eq:mask objective})
(\emph{Sub-Problem 1}), it suffers from ill-posedness and non-uniqueness
as outlined in \cite{Salucci 2018} when dealing with the design of
reflectarray surface currents \cite{Huang 2008}. Moreover, it is
worth pointing out that the design method used in \cite{Salucci 2018}
cannot be directly translated to the \emph{SPSS} case since it fits
a pattern matching objective instead of a {}``footprint pattern mask
constrained'' one (\ref{eq:mask objective}). Thus, a different solution
strategy inspired by the \emph{IPT} \cite{Rocca 2009b} is proposed
hereinafter. Towards this purpose, the {}``pattern'' feasible space\begin{equation}
\mathcal{F}\left\{ \left[\mathbf{E}^{FF}\left(\mathbf{r}\right)\right]^{*}\right\} \triangleq\left\{ \left[\mathbf{E}^{FF}\left(\mathbf{r}\right)\right]^{*}:\,\mathcal{L}\left(\mathbf{r}\right)\leq\left|\left[\mathbf{E}^{FF}\left(\mathbf{r}\right)\right]^{*}\right|^{2}\leq\mathcal{U}\left(\mathbf{r}\right);\mathbf{r}\in\Xi\right\} \label{eq:e feasible}\end{equation}
and the {}``current'' feasible space\begin{equation}
\mathcal{F}\left\{ \left[\mathbf{J}^{w}\left(\mathbf{r}\right)\right]^{*}\right\} \triangleq\left\{ \left[\mathbf{J}^{w}\left(\mathbf{r}\right)\right]^{*}:\,\left[\mathbf{J}^{w}\left(\mathbf{r}\right)\right]^{*}=C^{w}\exp\left[j\psi^{w}\left(\mathbf{r}\right)\right];\mathbf{r}\in\Omega\right\} \label{eq:j feasible}\end{equation}
are firstly defined, where $C^{w}$ and $\psi^{w}\left(\mathbf{r}\right)$
are the constant magnitude and the profile of the locally-controlled
phase of the $w$-th ($w=\left\{ e,m\right\} $) current component,
respectively. While different scenarios and assumptions can be accounted
for defining the feasibility spaces (\ref{eq:e feasible}) and (\ref{eq:j feasible}),
one should notice that the statement in (\ref{eq:j feasible}) implies
that the \emph{SPSS} unit cells do not to allow a control of the local
magnitude of the electric/magnetic currents. Subject to these assumptions,
the \emph{IPT}-based design of the \emph{SPSS} currents design (Fig.
2) is implemented according to the following iterative procedure ($h=1,...,H$
being the iteration index) \cite{Rocca 2009b}

\begin{itemize}
\item \noindent \emph{Initialization} ($h=0$) - The $w$-th ($w=\left\{ e,m\right\} $)
surface current is discretized\begin{equation}
\mathbf{J}_{h}^{w}\left(\mathbf{r}\right)=\sum_{p=1}^{P}\sum_{q=1}^{Q}\left[\left(J_{h}^{w}\right)_{x}^{pq}\widehat{\mathbf{x}}+\left(J_{h}^{w}\right)_{y}^{pq}\widehat{\mathbf{y}}\right]\Pi^{pq}\left(\mathbf{r}\right),\label{eq:current discretization}\end{equation}
the expansion coefficients, \{$\left(J_{h}^{w}\right)_{x}^{pq}$;
$p=1,..,P$; $q=1,...,Q$\} and \{$\left(J_{h}^{w}\right)_{y}^{pq}$;
$p=1,..,P$; $q=1,...,Q$\} being set to random values such that the
condition $\left\Vert \left(J_{h}^{w}\right)_{x}^{pq}\widehat{\mathbf{x}}+\left(J_{h}^{w}\right)_{y}^{pq}\widehat{\mathbf{y}}\right\Vert =C^{w}$
($p=1,..,P$; $q=1,...,Q$) holds true;
\item \noindent \emph{Pattern Computation} - The far-field pattern $\mathbf{E}_{h}^{FF}\left(\mathbf{r}\right)$
is evaluated in the coverage region $\Xi$ by substituting (\ref{eq:current discretization})
in (\ref{eq:far field});
\item \emph{Projection to Pattern Feasibility Space} - The projected pattern
$\widetilde{\mathbf{E}}_{h}^{FF}\left(\mathbf{r}\right)$ is obtained
by setting \begin{equation}
\widetilde{\mathbf{E}}_{h}^{FF}\left(\mathbf{r}\right)=\left\{ \begin{array}{ll}
\sqrt{\mathcal{U}\left(\mathbf{r}\right)} & \mathrm{if}\,\left|\mathbf{E}_{h}^{FF}\left(\mathbf{r}\right)\right|^{2}>\mathcal{U}\left(\mathbf{r}\right)\\
\sqrt{\mathcal{L}\left(\mathbf{r}\right)} & \mathrm{if}\,\left|\mathbf{E}_{h}^{FF}\left(\mathbf{r}\right)\right|^{2}<\mathcal{L}\left(\mathbf{r}\right)\\
\mathbf{E}_{h}^{FF}\left(\mathbf{r}\right) & \mathrm{otherwise};\end{array}\right.\label{eq:projection}\end{equation}

\item \emph{Convergence Check} - The algorithm is terminated by returning
the ideal/reference $w$-th ($w=\left\{ e,m\right\} $) surface current,
$\left[\mathbf{J}^{w}\left(\mathbf{r}\right)\right]^{*}=\mathbf{J}_{h}^{w}\left(\mathbf{r}\right)$,
if $h=H$ or the \emph{pattern matching index}, $\mathcal{X}_{h}$,\begin{equation}
\mathcal{X}_{h}=\frac{\int_{\Xi}\left|\widetilde{\mathbf{E}}_{h}^{FF}\left(\mathbf{r}\right)-\mathbf{E}_{h}^{FF}\left(\mathbf{r}\right)\right|^{2}\mathrm{d}\mathbf{r}}{\int_{\Xi}\left|\mathbf{E}_{h}^{FF}\left(\mathbf{r}\right)\right|^{2}\mathrm{d}\mathbf{r}}\label{eq:pattern convergence}\end{equation}
satisfies the condition $\mathcal{X}_{h}\le\mathcal{X}^{*}$, $\mathcal{X}^{*}$
being a user-chosen convergence threshold;
\item \emph{Computation of Minimum Norm Currents} - Compute the \emph{minimum
norm} component of the $w$-th ($w=\left\{ e,m\right\} $) surface
current, $\left[\mathbf{J}_{h}^{w}\left(\mathbf{r}\right)\right]^{MN}$,
by solving (\ref{eq:far field}) with respect to the currents. Towards
this end, the method based on the truncated singular value decomposition,
detailed in \cite{Salucci 2018}, is applied;
\item \emph{Projection to Current Feasibility Space} - Update the iteration
index ($h\leftarrow h+1$) and evaluate the $w$-th ($w=\left\{ e,m\right\} $)
projected surface current $\mathbf{J}_{h}^{w}\left(\mathbf{r}\right)$

\begin{equation}
\mathbf{J}_{h}^{w}\left(\mathbf{r}\right)=C^{w}\sum_{p=1}^{P}\sum_{q=1}^{Q}\frac{\left[\left(J_{h-1}^{w}\right)_{x}^{pq}\right]^{MN}\widehat{\mathbf{x}}+\left[\left(J_{h-1}^{w}\right)_{y}^{pq}\right]^{MN}\widehat{\mathbf{y}}}{\left\Vert \left[\left(J_{h-1}^{w}\right)_{x}^{pq}\right]^{MN}\widehat{\mathbf{x}}+\left[\left(J_{h-1}^{w}\right)_{y}^{pq}\right]^{MN}\widehat{\mathbf{y}}\right\Vert }\Pi^{pq}\left(\mathbf{r}\right).\label{eq:correnti proiettate}\end{equation}
Restart process from the {}``\emph{Pattern Computation}'' step.

\end{itemize}
\noindent Once the reference surface currents, \{$\left[\mathbf{J}^{w}\left(\mathbf{r}\right)\right]^{*}$;
$w=\left\{ e,m\right\} $\}, have been found, the \emph{Sub-problem
2} is then addressed by solving (\ref{eq:current matching}). More
in detail, an iterative \emph{SbD}-based strategy inspired by \cite{Oliveri 2020}
is customized to the problem at hand by implementing the following
blocks of the functional flowchart in Fig. 2: (\emph{i}) the {}``Solution
Space Exploration (\emph{SSE})'' block aimed at optimizing the \emph{SPSS}
descriptors by defining a succession of $S$ iterations ($s$ being
the \emph{SbD} iteration index, $s=1,...,S$) where $G$ trial solutions,
\{$\mathcal{D}_{g}^{\left(s\right)}$; $g=1,...,G$\} (i.e., the population
$\mathcal{D}^{\left(s\right)}$), $\mathcal{D}_{g}^{\left(s\right)}\triangleq\left\{ \left.\mathbf{d}_{pq}\right|_{g}^{\left(s\right)};\, p=1,...,P;\, q=1,...,Q\right\} $
being the $g$-th one at the $s$-th iteration, evolve towards the
global solution $\mathcal{D}^{opt}$ (\ref{eq:current matching}).
Because of the non-linear function to be optimized and the ill-posed
nature of the problem at hand, a global search mechanism based on
the \emph{Particle Swarm Optimizer} \cite{Rocca 2009} has been chosen
to update/evolve the population of trial solutions at each $s$-th
($s=1,...,S$) step, $\mathcal{D}^{\left(s\right)}$; (\emph{ii})
the {}``Cost Function (\emph{CF})'' evaluation block that implements
the discretized version of (\ref{eq:current matching}); (\emph{iii})
the {}``Surface Current Evaluation (\emph{SCE})'' block that employs
(\ref{eq:GSTC Je}) and (\ref{eq:GSTC Jm}) to determine $\mathbf{J}^{w}\left(\mathbf{r}\right)$
starting from $\mathbf{B}^{w}\left(\mathbf{r}\right)$, $w=\left\{ e,m\right\} $;
(\emph{iv}) the {}``Polarization Surface Densities Evaluation (\emph{PSDE})''
block that implements (\ref{eq:electric polariz}) and (\ref{eq:magnetic polar})
to yield the $w$-th ($w=\left\{ e,m\right\} $) polarization surface
density, $\mathbf{B}^{w}\left(\mathbf{r}\right)$; (\emph{v}) the
{}``Local Susceptibility Dyadics Digital Twin (\emph{LSDDT})'' block
devoted to determine $\overline{\overline{\chi}}\left(\left.\mathbf{d}_{pq}\right|_{g}^{\left(s\right)}\right)$
and $\overline{\overline{\xi}}\left(\left.\mathbf{d}_{pq}\right|_{g}^{\left(s\right)}\right)$
to be used in the \emph{PSDE} block to compute the polarization surface
densities at each $s$-th ($s=1,...,S$) iteration for each $g$-th
($g=1,...,G$) guess solution in each ($p$, $q$)-th ($p=1,...,P$;
$q=1,...,Q$) unit cell of the \emph{SPSS}. 

\noindent As for this latter and analogously to the unit cells of
reflectarrays \cite{Oliveri 2020}\cite{Salucci 2018c}, the full-wave
evaluation of each susceptibility tensor set, $\overline{\overline{\chi}}\left(\left.\mathbf{d}_{pq}\right|_{g}^{\left(s\right)}\right)$
and $\overline{\overline{\xi}}\left(\left.\mathbf{d}_{pq}\right|_{g}^{\left(s\right)}\right)$,
generated in the \emph{SbD} iterative process turns out computationally
unfeasible since this would require the numerical modelling and the
full-wave solution of $P\times Q\times G\times S$ \emph{SPSS}s. Therefore,
the dyadics $\overline{\overline{\chi}}\left(\mathbf{d}\right)$ and
$\overline{\overline{\xi}}\left(\mathbf{d}\right)$ are approximated
with their surrogates $\overline{\overline{\chi}}'\left(\mathbf{d}\right)$
and $\overline{\overline{\xi}}'\left(\mathbf{d}\right)$ defined by
a trained \emph{Digital Twin} (\emph{DT}), which is implemented according
to a statistical learning approach based on the \emph{Ordinary Kriging}
(\emph{OK}) method \cite{Oliveri 2020}\cite{Salucci 2018c}. This
choice is related to the effectiveness of the \emph{OK} in defining
accurate and reliable surrogate models of wave manipulating devices
\cite{Oliveri 2020}\cite{Salucci 2018c}. On the other hand, the
reader should consider that here, unlike the reflectarray case \cite{Oliveri 2020}\cite{Salucci 2018c},
the \emph{DT} has to predict the local susceptibility tensors rather
than the local reflection coefficient. This means that $6$ complex
coefficients (i.e., diagonal entries of $\overline{\overline{\chi}}\left(\mathbf{d}\right)$
and $\overline{\overline{\xi}}\left(\mathbf{d}\right)$) must be taken
into account instead of $4$ terms (i.e., the $2\times2$ entries
of the reflection matrix \cite{Oliveri 2020}\cite{Salucci 2018c}),
but also that the \emph{DT} of a \emph{SPSS} can neglect the incidence
angle of the illuminating field since the susceptibility tensor, unlike
the reflection coefficients \cite{Yang 2019}, does not depend on
it.

\section{\noindent Numerical Results\label{sec:Numerical-Analysis-and}}

\noindent This section is aimed at illustrating the \emph{IPT-SbD}
design process and at numerically assessing its effectiveness in synthesizing
holographic \emph{SPSS}s suitable for footprint pattern shaping. Besides
the value of the pattern matching index, $\mathcal{X}$ (\ref{eq:pattern convergence}),
of the \emph{SPSS} final layout (i.e., $\mathcal{X}^{SPSS}$), the
accuracy of each step of the synthesis process has been also {}``quantified''
by computing the \emph{reference pattern matching} $\mathcal{X}^{IPT}$
($\mathcal{X}^{IPT}\triangleq\mathcal{X}_{H}$ - \emph{Sub-Problem
1}) and the \emph{surface current fidelity index} $\upsilon^{SbD}$
($\upsilon^{SbD}\triangleq\upsilon\left(\left.\mathbf{J}^{w}\left(\mathbf{r}\right)\right\rfloor _{s=S};\,\left[\mathbf{J}^{w}\left(\mathbf{r}\right)\right]^{*}\right)$
- \emph{Sub-Problem 2}). In the numerical analysis, different \emph{SPSS}
apertures and target footprint masks have been considered by assuming
a benchmark \emph{SEE} scenario where a base station illuminates from
$\left(\theta^{inc},\varphi^{inc}\right)=\left(20,105\right)$ {[}deg{]}
the smart skin with a linearly-polarized plane wave having a slant
$+45$ {[}deg{]} polarization at $f=30$ {[}GHz{]}.

\noindent As for the metasurface unit cell, a square metallic patch
with periodicity $\delta_{x}=\delta_{y}=5.0\times10^{-3}$ {[}m{]}
printed on a single-layer substrate (Rogers 3003 dielectric with thickness
$\tau=5.08\times10^{-4}$ {[}m{]}) has been used ($L=1$) and modeled
in \emph{HFSS} \cite{HFSS 2019} for generating/training the \emph{LSDDT}
block \cite{Oliveri 2020}\cite{Salucci 2018c}. Such a simple structure
has been chosen to highlight the potentials of the \emph{IPT-SbD}
strategy even when dealing with elementary unit cells. As for the
\emph{IPT-SbD} parametric configuration, the following setup has been
chosen according to the guidelines in \cite{Oliveri 2020}: $H=10^{3}$,
$\mathcal{X}^{*}=10^{-4}$, $S=10^{4}$, and $G=10$.

\noindent The first numerical experiment deals with a $P\times Q=200\times200$
holographic \emph{SPSS} with an $1\times1$ {[}m{]} support $\Omega$
(Fig. 1) located at the position $\left(x',y',z'\right)=\left(0,0,15\right)$
{[}m{]} in the global coordinate system, $\mathbf{r}'=\left(x',y',z'\right)$.
Moreover, the upper and the lower masks have been defined so that
the skin reflects a constant-power square footprint in the coverage
region $\Xi$ of lateral size $10$ {[}m{]} centered at $\left(x',y',z'\right)=\left(-25,25,0\right)$
{[}m{]} {[}{}``Square Footprint'' - Fig. 3(\emph{a}){]}, while a
$-30$ {[}dB{]} footprint power reduction has been enforced outside
$\Xi$ in the observation region $\Theta$ of extension $120\times60$
{[}$\textnormal{m}{}^{2}${]}. According to the proposed design approach
(Fig. 2), the synthesis of the $w$-th ($w=\left\{ e,m\right\} $)
\emph{ideal} surface current, $\left[\mathbf{J}^{w}\left(\mathbf{r}\right)\right]^{*}$,
has been carried out by solving the associated \emph{IS} problem through
the \emph{IPT}-based iterative procedure. The evolution of the \emph{IPT}
cost function during the iterative process, $\mathcal{X}_{h}$ ($h=1,...,H$)
{[}Fig. 3(\emph{b}){]}, shows that there is a quick minimization {[}i.e.,
$\frac{\mathcal{X}_{h}}{\mathcal{X}_{0}}<10^{-3}$ when $h\geq25$
- Fig. 3(\emph{b}){]} and a convergence to a solution with a very
small mismatch from the target footprint pattern, $\mathcal{X}_{H}=6.44\times10^{-4}$
(Tab. I) in less than $4$ minutes%
\footnote{\noindent For the sake of fairness, all the computation times refer
to non-optimized MATLAB implementations executed on a single-core
laptop running at $1.60$ GHz.%
} (Tab. I) thanks to the exploitation of a fast Fourier transform within
the \emph{IPT} loop despite the huge number of unknowns (i.e., $N_{\mathcal{D}}=4.0\times10^{4}$).
For illustrative purposes, the phase of the dominant component (i.e.,
slant $+45$ {[}deg{]} polarization) of the synthesized ideal current
is reported in Fig. 4(\emph{a}). Concerning the second step (\emph{Sub-Problem
2}) aimed at determining the \emph{SPSS} layout that supports the
\emph{IPT}-computed reference currents, the \emph{SbD} optimization
process quickly ($\Delta t^{SbD}<10$ {[}s{]} - Tab. I) yields, thanks
to an accurate matching (i.e., $\upsilon^{SbD}=2.05\times10^{-1}$
- Tab. I) with the reference current {[}Fig. 4(\emph{b}) vs. Fig.
4(\emph{a}){]}, a final layout {[}Fig. 4(\emph{c}){]} that faithfully
fulfils the mask requirements (i.e., $\mathcal{X}^{SPSS}=1.08\times10^{-3}$)
as pictorially confirmed by the plot of the radiated footprint pattern
within the observation region {[}Fig. 5(\emph{a}) vs. Fig. 3(\emph{a}){]}.
For completeness, the angular power distribution is reported in Fig.
5(\emph{b}) to point out the {}``focusing'' skills of the synthesized
\emph{SPSS} or, in other words, the ability of such a holographic
metasurface to compensate the angular beam distortion caused by the
position and the orientation of the coverage region with respect to
the smart skin and the incident wave.

\noindent The feasibility of the \emph{SPSS} synthesis is checked
next against the more challenging {}``Checkerboard'' footprint mask
{[}Fig. 6(\emph{a}){]}. Although the problem at hand features a more
complex target footprint, the arising holographic arrangement {[}Fig.
6(\emph{b}){]} fits the radiation constraints {[}Fig. 7(\emph{a})
vs. Fig. 6(\emph{a}){]} with an effective angular control of the radiated
power {[}Fig. 7(\emph{b}){]}. It is also interesting to note that
the greater complexity of the pattern mask {[}Fig. 6(\emph{a}) vs.
Fig. 3(\emph{a}){]} impacts neither on the \emph{CPU} time for synthesis
process {[}$\left.\Delta t^{IPT}\right\rfloor _{Checkerboard}=2.37\times10^{2}$
{[}s{]} vs. $\left.\Delta t^{IPT}\right\rfloor _{Square}=2.31\times10^{2}$
{[}s{]} and $\left.\Delta t^{SbD}\right\rfloor _{Checkerboard}=9.08$
{[}s{]} vs. $\left.\Delta t^{SbD}\right\rfloor _{Square}=9.30$ {[}s{]}
- Tab. I{]} nor on the convergence of the two-step synthesis as quantitatively
confirmed by the values of the currents and pattern matching indexes
in Tab. I (e.g., $\left.\mathcal{X}^{SPSS}\right\rfloor _{Checkerboard}=7.78\times10^{-4}$
vs. $\left.\mathcal{X}^{SPSS}\right\rfloor _{Square}=1.08\times10^{-3}$
and $\left.\upsilon^{SbD}\right\rfloor _{Checkerboard}=2.04\times10^{-1}$
vs. $\left.\upsilon^{SbD}\right\rfloor _{Square}=2.05\times10^{-1}$).

\noindent The possibility to simultaneously cover a wider region (i.e.,
$\Xi$ is a rectangle of $20\times80$ {[}m{]} modeling a short street
in front of the smart skin) with a locally-complex footprint is addressed
next by dealing with the {}``\emph{IEEE}'' shape in Fig. 8(\emph{a}).
Figure 8(\emph{b}) shows the behavior of the \emph{IPT} cost function
$\mathcal{X}_{h}$ during the iterative optimization of the currents
distribution towards the reference one ($\mathcal{X}^{IPT}=3.65\times10^{-3}$)
then approximated ($\upsilon^{SbD}=2.06\times10^{-1}$) by the \emph{SbD}
layout in Fig. 9(\emph{a}) that radiates the well controlled footprint
in Fig. 9(\emph{b}) ($\mathcal{X}^{SPSS}=4.84\times10^{-3}$ vs. $\mathcal{X}^{IPT}=3.65\times10^{-3}$).
As it can be observed, the synthesized \emph{SPSS} not only fits the
footprint mask, but also compensates the path loss to generate uniform
levels of power over within the observation region $\Theta$ at considerably
different distances from the smart skin {[}Fig. 9(\emph{b}){]}.

\noindent Finally, the last experiment is devoted to assess the proposed
design approach as well as its dependence on the \emph{SPSS} aperture
when dealing with advanced beamforming tasks involving detailed footprint
shapes. Towards this end, the {}``\emph{ELEDIA}'' mask {[}Fig. 8(\emph{c}){]}
has been considered and the \emph{SPSS} design has been carried out
by varying its support $\Omega$ from $P\times Q=25\times25$ {[}i.e.,
$N_{\mathcal{D}}=625$ - Fig. 10(\emph{a}){]} up to $P\times Q=400\times400$
unit cells {[}i.e., $N_{\mathcal{D}}=1.6\times10^{5}$ - Fig. 10(\emph{f}){]}
and the corresponding footprints are shown in Figs. 11(\emph{a})-11(\emph{f}).
For completeness, Figure 12 gives the plots of the matching indexes.
From these results, one can infer the following outcomes: (a) unless
the smallest apertures (i.e., $P\times Q=25\times25$), the proposed
\emph{IPT-SbD} approach can handle complex footprints {[}see Figs.
11(\emph{c})-11(\emph{f}) vs. Fig. 8(\emph{c}){]}; (\emph{b}) as expected,
it profitably leverages the increased number of descriptors of wider
apertures to improve the beamforming accuracy as quantitatively confirmed
by the behavior of $\mathcal{X}^{SPSS}=\mathcal{X}^{SbD}$, $\upsilon^{SbD}$
and $\mathcal{X}^{IPT}$ in Fig. 12(\emph{b}) and Tab. I as well as
by the evolution of the \emph{IPT} process versus the iteration number
$h$ ($h=1,...,H$) {[}Fig. 12(\emph{a}){]}; (\emph{c}) the entire
synthesis process turns out to be extremely efficient whatever the
number of \emph{DoF}s and pattern footprint. As a representative example,
the reader can consider that when $P\times Q=400\times400$, the whole
CPU-time is $\Delta t^{IPT}+\Delta t^{SbD}<18$ {[}min{]}.

\section{\noindent Conclusions\label{sec:Conclusions-and-Remarks}}

\noindent The possibility to efficiently and effectively synthesize
inexpensive smart \emph{EM} skins supporting advanced beamforming
capabilities has been addressed. More specifically, the design of
passive/static smart skins with enhanced wave manipulation capabilities
has been formulated within the \emph{GSTC} theoretical framework by
exploiting an \emph{IS} formulation. An integrated synthesis procedure
has been then proposed that combines a mask-constrained isophoric
source design based on the \emph{IPT} and a \emph{SbD}-driven optimization
for determining the \emph{SPSS} layout fitting user-defined beam-pattern
requirements. The feasibility of suitable cheap and passive wave manipulation
holographic metasurfaces has been assessed as well as the effectiveness
of the proposed synthesis approach against different footprint targets,
skin dimensions, and coverage regions. The outcomes from such a numerical
validation have confirmed that structurally simple yet high-performance
holographic metasurfaces can be yielded {[}e.g., Fig. 10{]} with the
proposed \emph{SPSS} design process that efficiently handles large
apertures, as well (Tab. I). Future works, beyond the scope of this
paper, will be devoted to extend such a design technique to multi-function
and reconfigurable smart \emph{EM} skins as well as to analyze its
potentials when exploiting more complex unit cells and non-uniform/unconventional
arrangements of the {}``re-radiating'' elements \cite{Rocca 2016}.

\section*{\noindent Acknowledgements}

\noindent This work benefited from the networking activities carried
out within the Project {}``Cloaking Metasurfaces for a New Generation
of Intelligent Antenna Systems (MANTLES)'' (Grant No. 2017BHFZKH)
funded by the Italian Ministry of Education, University, and Research
under the PRIN2017 Program (CUP: E64I19000560001). Moreover, it benefited
from the networking activities carried out within the Project {}``SPEED''
(Grant No. 61721001) funded by National Science Foundation of China
under the Chang-Jiang Visiting Professorship Program, the Project
'Inversion Design Method of Structural Factors of Conformal Load-bearing
Antenna Structure based on Desired EM Performance Interval' (Grant
no. 2017HZJXSZ) funded by the National Natural Science Foundation
of China, and the Project 'Research on Uncertainty Factors and Propagation
Mechanism of Conformal Loab-bearing Antenna Structure' (Grant No.
2021JZD-003) funded by the Department of Science and Technology of
Shaanxi Province within the Program Natural Science Basic Research
Plan in Shaanxi Province. A. Massa wishes to thank E. Vico for her
never-ending inspiration, support, guidance, and help.

\newpage
\section*{FIGURE CAPTIONS}

\begin{itemize}
\item \textbf{Figure 1.} \emph{Problem geometry}. Sketch of the smart \emph{EM}
skin scenario.
\item \textbf{Figure 2.} \emph{SPSS Design Approach}. Flowchart of the \emph{IPT-SbD}
holographic metasurface synthesis process.
\item \textbf{Figure 3.} \emph{Numerical Validation} (\emph{{}``Square''
Footprint}, $P=Q=200$) - Plot of (\emph{a}) the footprint pattern
mask {[}$\mathcal{U}\left(\mathbf{r}'\right)$; $\mathbf{r}'\in\Theta${]}
and (\emph{b}) evolution of the \emph{IPT} cost function versus the
iteration index, $h$ ($h=1,...,H$).
\item \textbf{Figure 4.} \emph{Numerical Validation} (\emph{{}``Square''
Footprint}, $P=Q=200$) - Plot of the phase distribution of (\emph{a})
the \emph{IPT} reference/ideal current along with (\emph{b}) that
generated \emph{}by the synthesized \emph{SPSS} layout (c).
\item \textbf{Figure 5.} \emph{Numerical Validation} (\emph{{}``Square''
Footprint}, $P=Q=200$) - Plots of the radiated (\emph{a}) footprint
pattern within the observation region $\Theta$ and (\emph{b}) angular
power distribution.
\item \textbf{Figure 6.} \emph{Numerical Validation} (\emph{{}``Checkerboard''
Footprint}, $P=Q=200$) - Plot of (\emph{a}) the footprint pattern
mask {[}$\mathcal{U}\left(\mathbf{r}'\right)$; $\mathbf{r}'\in\Theta${]}
and (\emph{b}) layout of the synthesized \emph{SPSS}.
\item \textbf{Figure 7.} \emph{Numerical Validation} (\emph{{}``Checkerboard''
Footprint}, $P=Q=200$) - Plots of the radiated (\emph{a}) footprint
pattern within the observation region $\Theta$ and (\emph{b}) angular
power distribution.
\item \textbf{Figure 8.} \emph{Numerical Validation} ($P=Q=200$) - Plots
of (\emph{a}) the {}``\emph{IEEE}'' and (\emph{c}) the {}``\emph{ELEDIA}''
footprint pattern mask {[}$\mathcal{U}\left(\mathbf{r}'\right)$;
$\mathbf{r}'\in\Theta${]} along with the (\emph{b}) evolution of
the \emph{IPT} cost function versus the iteration index, $h$ ($h=1,...,H$).
\item \textbf{Figure 9.} \emph{Numerical Validation} (\emph{{}``IEEE''
Footprint}, $P=Q=200$) - Plots of (\emph{b}) the \emph{SPSS} layout
and of (\emph{a}) the corresponding footprint pattern within the observation
region $\Theta$.
\item \textbf{Figure 10.} \emph{Numerical Validation} (\emph{{}``ELEDIA''
Footprint}) - Layouts of the synthesized \emph{SPSS} when (\emph{a})
$P=Q=25$, (\emph{b}) $P=Q=50$, (\emph{c}) $P=Q=100$, (\emph{d})
$P=Q=150$, (\emph{e}) $P=Q=200$, and (\emph{f}) $P=Q=400$.
\item \textbf{Figure 11.} \emph{Numerical Validation} (\emph{{}``ELEDIA''
Footprint}) - Footprint patterns radiated in the observation region
$\Theta$ by the \emph{SPSS} with (\emph{a}) $P=Q=25$ {[}Fig. 10(\emph{a}){]},
(\emph{b}) $P=Q=50$ {[}Fig. 10(\emph{b}){]}, (\emph{c}) $P=Q=100$
{[}Fig. 10(\emph{c}){]}, (\emph{d}) $P=Q=150$ {[}Fig. 10(\emph{d}){]},
(\emph{e}) $P=Q=200$ {[}Fig. 10(\emph{e}){]}, and (\emph{f}) $P=Q=400$
{[}Fig. 10(\emph{f}){]} unit cells.
\item \textbf{Figure 12.} \emph{Numerical Validation} (\emph{{}``ELEDIA''
Footprint}) - Plot of (\emph{a}) the evolution of the \emph{IPT} cost
function versus the iteration index, $h$ ($h=1,...,H$) and of (\emph{b})
the matching indexes ($\mathcal{X}^{SPSS}=\mathcal{X}^{SbD}$, $\upsilon^{SbD}$,
and $\mathcal{X}^{IPT}$) versus the \emph{SPSS} size.
\end{itemize}

\section*{TABLE CAPTIONS}

\begin{itemize}
\item \textbf{Table I.} \emph{Numerical Validation}. Matching and computational
indexes.
\end{itemize}
~

\newpage
\begin{center}~\vfill\end{center}

\begin{center}\includegraphics[%
  clip,
  width=0.95\columnwidth,
  keepaspectratio]{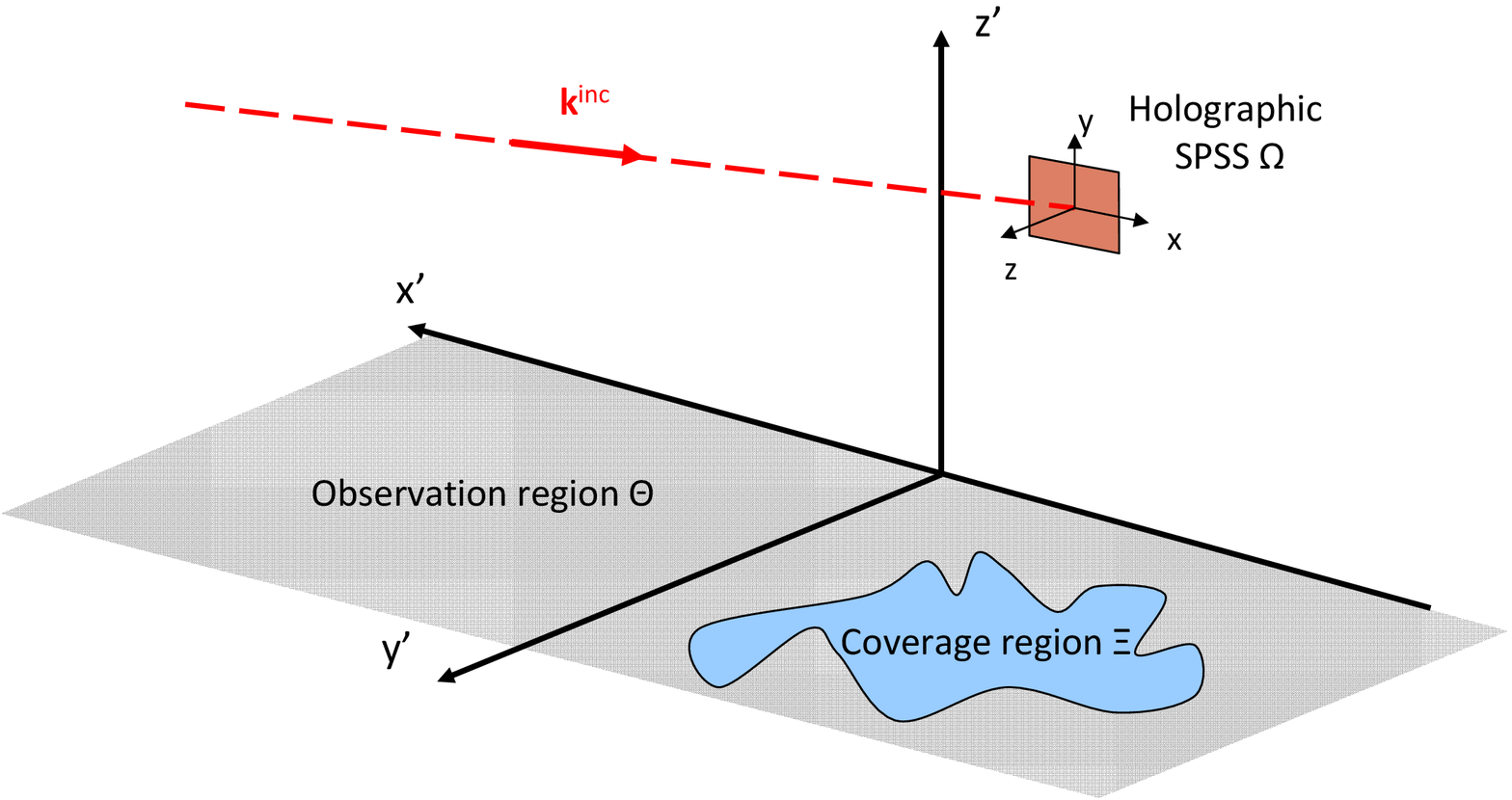}\end{center}

\begin{center}~\vfill\end{center}

\begin{center}\textbf{Fig. 1 - G. Oliveri et} \textbf{\emph{al.}}\textbf{,}
{}``Holographic Smart \emph{EM} Skins ...''\end{center}

\newpage
\begin{center}~\vfill\end{center}

\begin{center}\includegraphics[%
  clip,
  width=0.95\columnwidth,
  keepaspectratio]{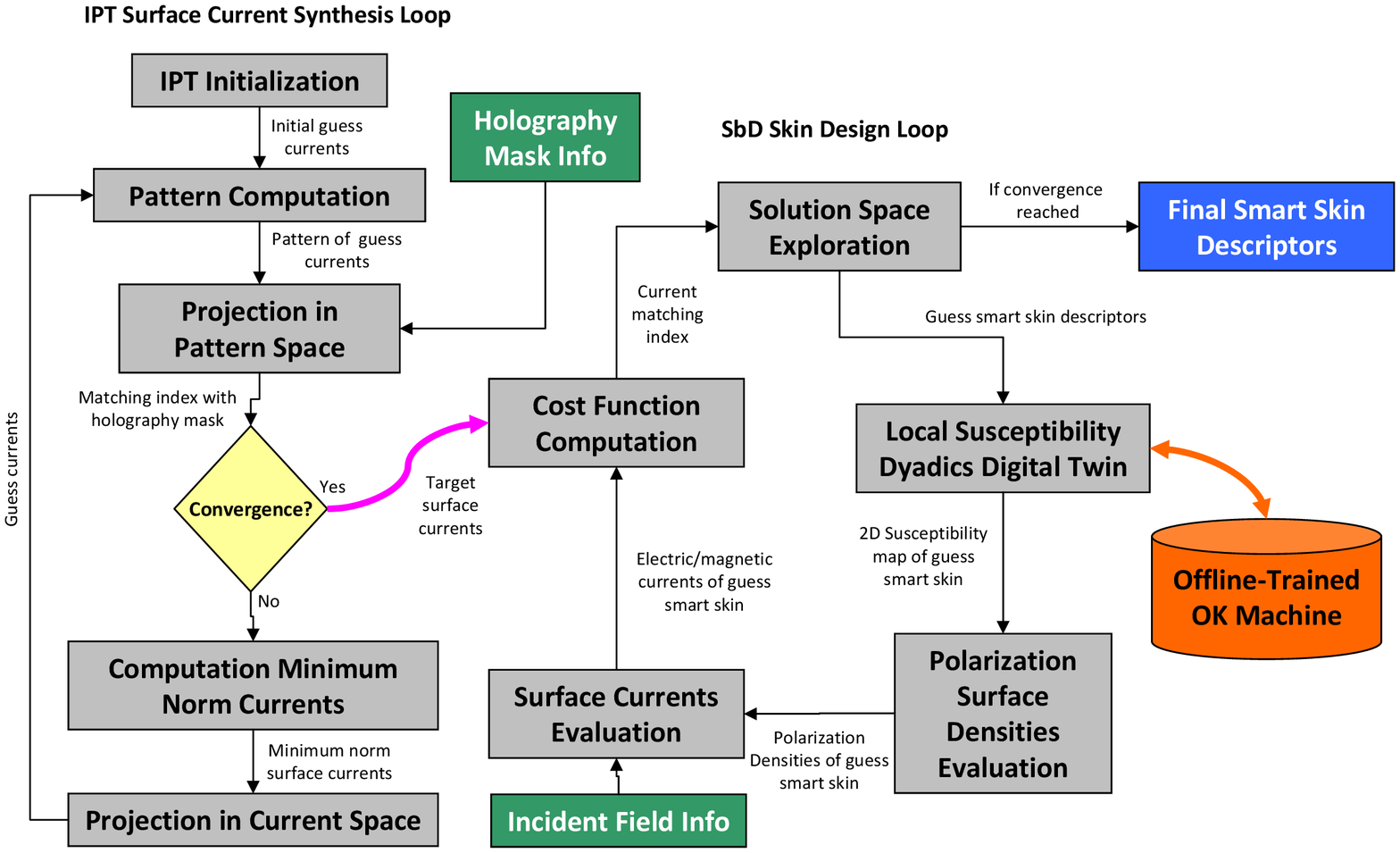}\end{center}

\begin{center}~\vfill\end{center}

\begin{center}\textbf{Fig. 2 - G. Oliveri et} \textbf{\emph{al.}}\textbf{,}
{}``Holographic Smart \emph{EM} Skins ...''\end{center}
\newpage

\begin{center}~\vfill\end{center}

\begin{center}\begin{tabular}{c}
\includegraphics[%
  clip,
  width=0.95\columnwidth,
  keepaspectratio]{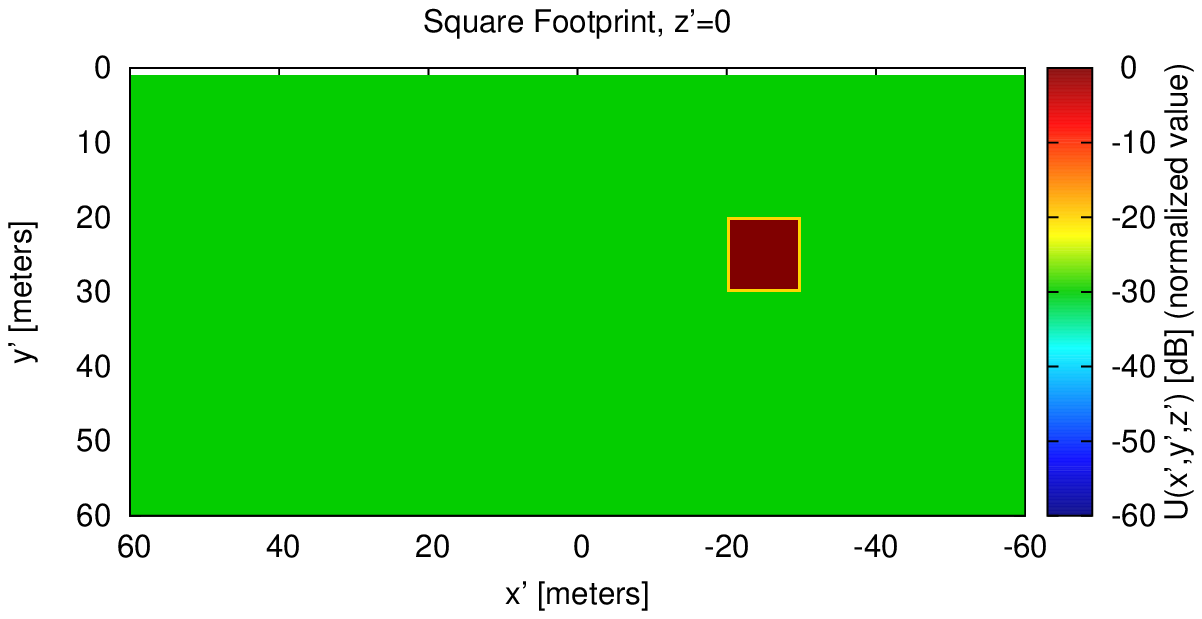}\tabularnewline
(\emph{a})\tabularnewline
\includegraphics[%
  clip,
  width=0.95\columnwidth,
  keepaspectratio]{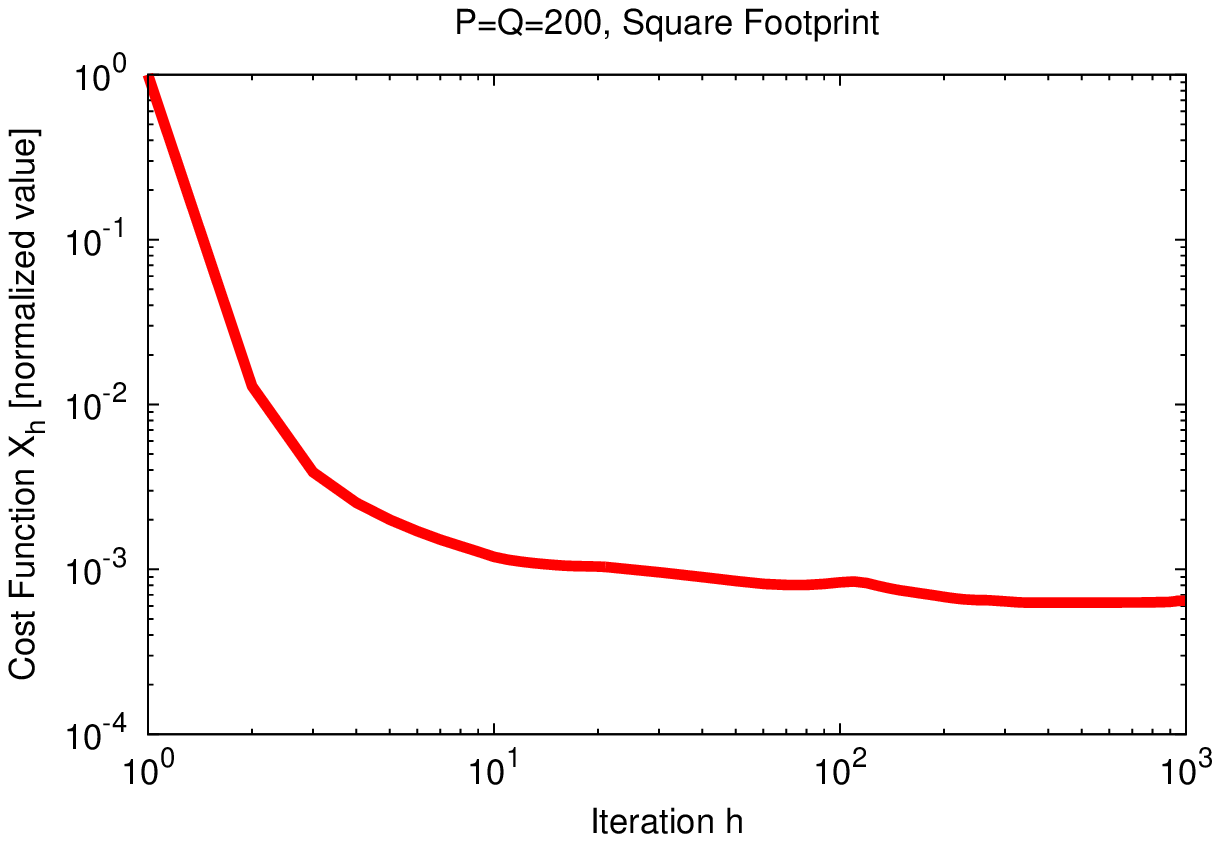}\tabularnewline
(\emph{b})\tabularnewline
\end{tabular}\end{center}

\begin{center}\vfill\end{center}

\begin{center}\textbf{Fig. 3 - G. Oliveri et} \textbf{\emph{al.}}\textbf{,}
{}``Holographic Smart \emph{EM} Skins ...''\end{center}
\newpage

\begin{center}\begin{tabular}{c}
\includegraphics[%
  clip,
  width=0.50\columnwidth,
  keepaspectratio]{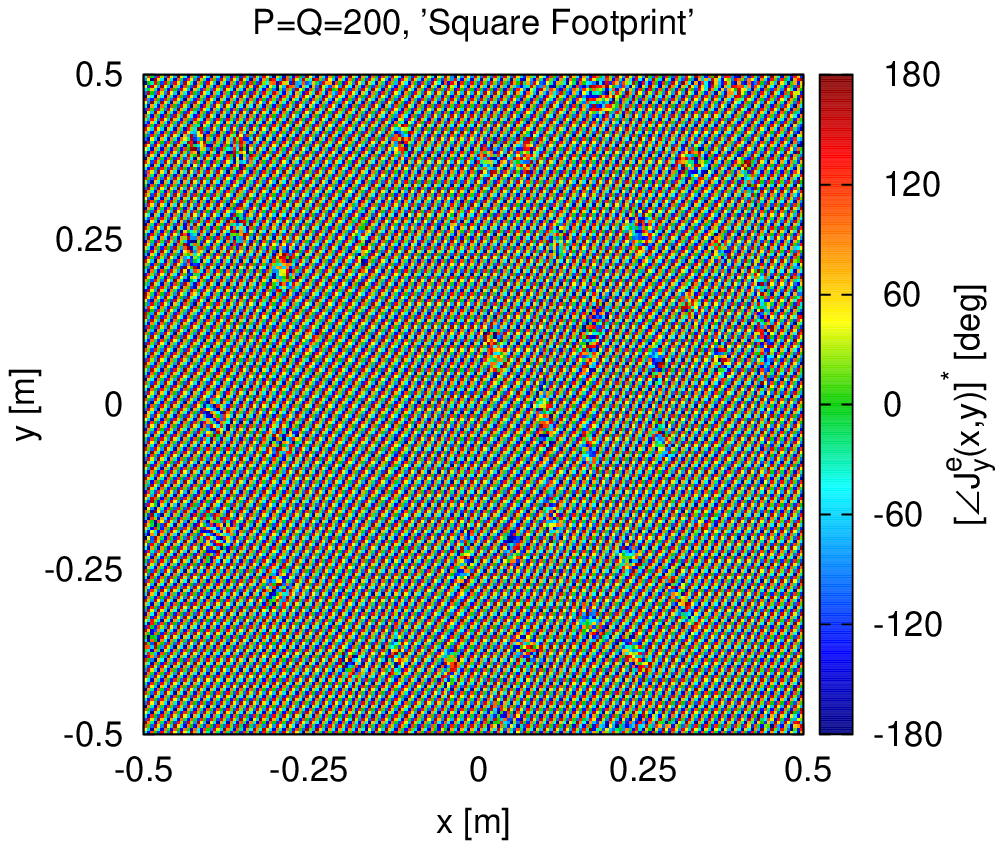}\tabularnewline
(\emph{a})\tabularnewline
\includegraphics[%
  clip,
  width=0.50\columnwidth,
  keepaspectratio]{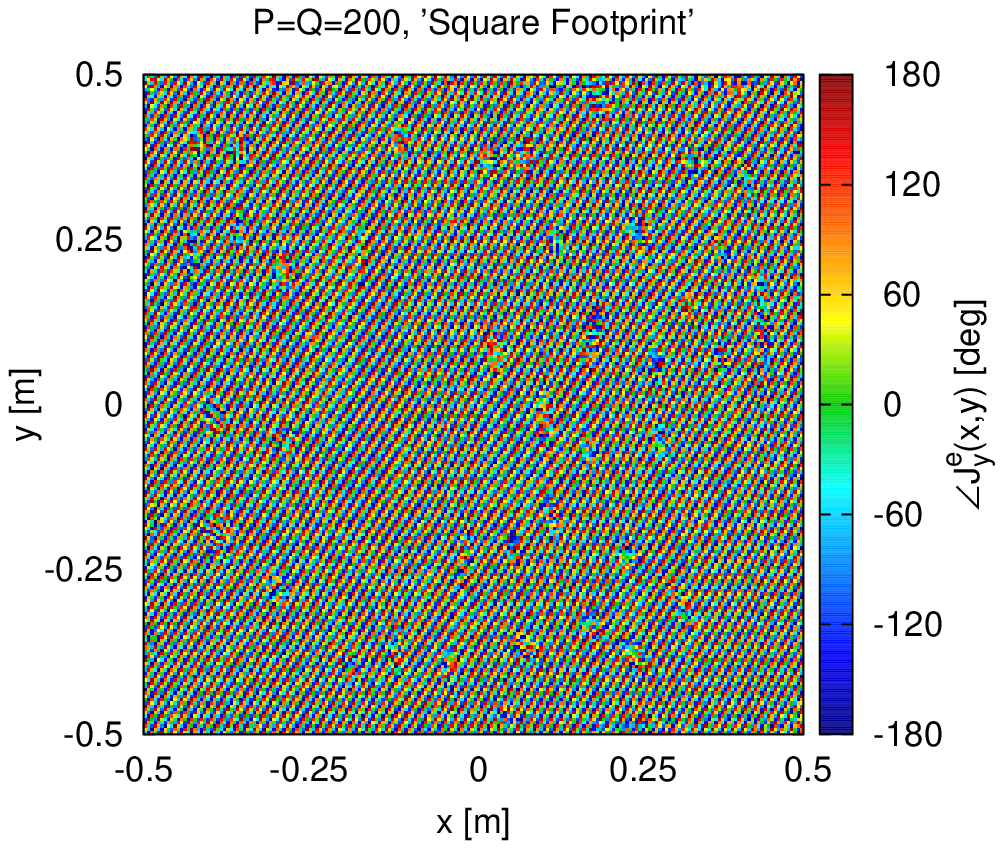}\tabularnewline
(\emph{b})\tabularnewline
\includegraphics[%
  clip,
  width=0.45\columnwidth,
  keepaspectratio]{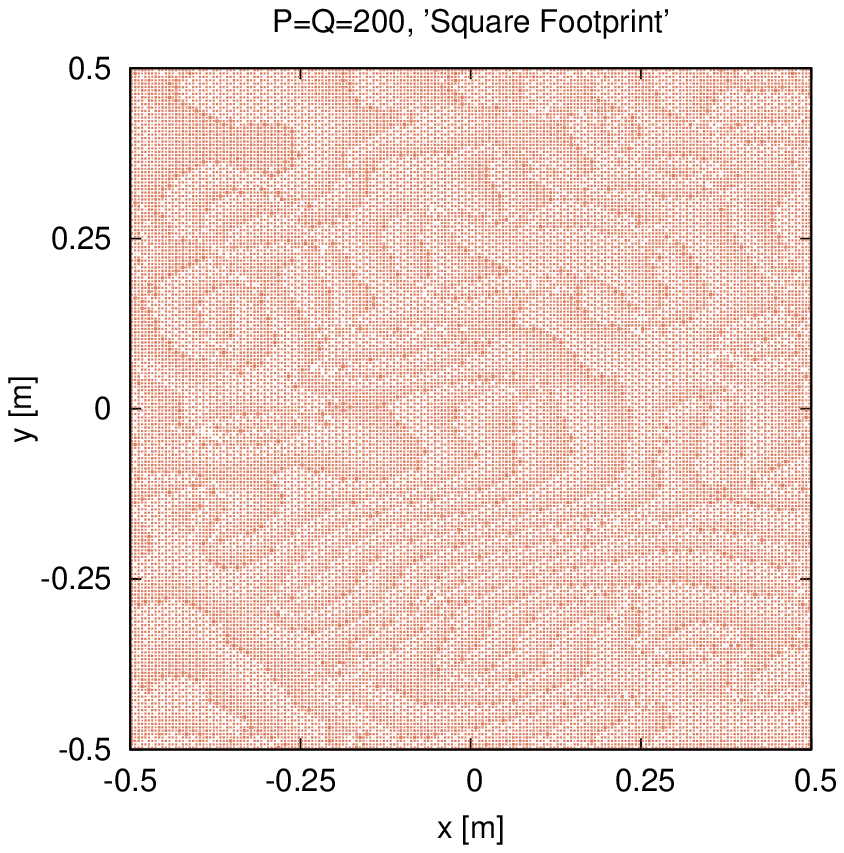}\tabularnewline
(\emph{c})\tabularnewline
\end{tabular}\end{center}

\begin{center}\textbf{Fig. 4 - G. Oliveri et} \textbf{\emph{al.}}\textbf{,}
{}``Holographic Smart \emph{EM} Skins ...''\end{center}
\newpage

\begin{center}\begin{tabular}{c}
\includegraphics[%
  clip,
  width=0.95\columnwidth,
  keepaspectratio]{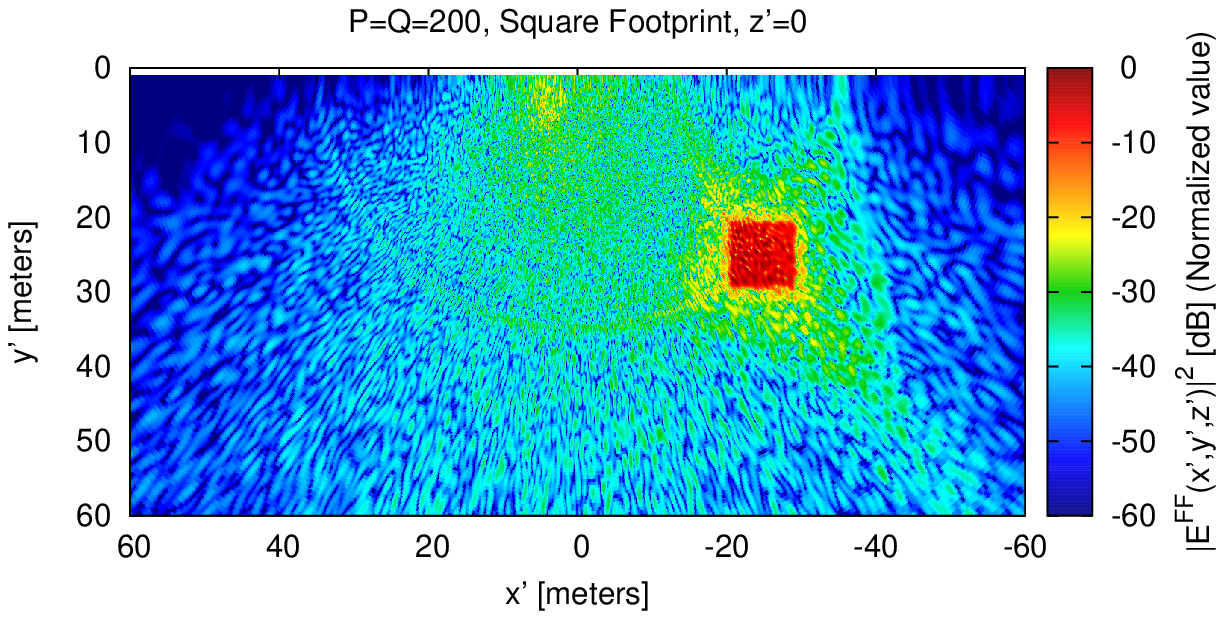}\tabularnewline
(\emph{a})\tabularnewline
\tabularnewline
\includegraphics[%
  clip,
  width=0.90\columnwidth,
  keepaspectratio]{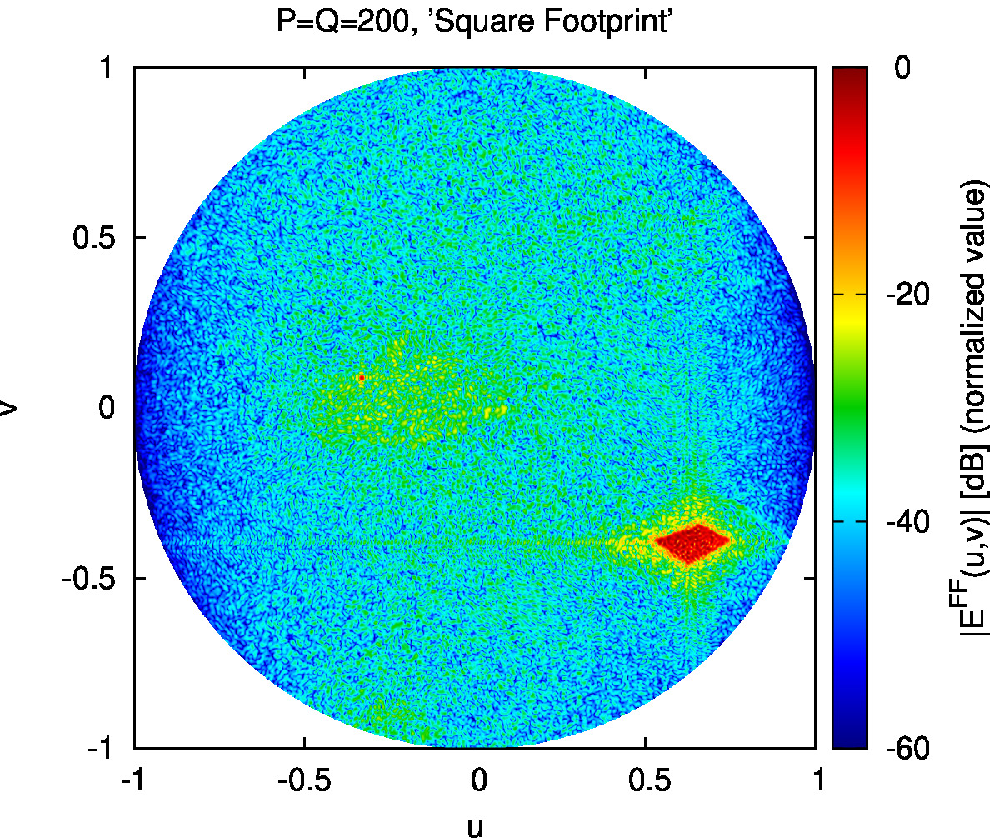}\tabularnewline
(\emph{b})\tabularnewline
\end{tabular}\end{center}

\begin{center}\vfill\end{center}

\begin{center}\textbf{Fig. 5 - G. Oliveri et} \textbf{\emph{al.}}\textbf{,}
{}``Holographic Smart \emph{EM} Skins ...''\end{center}
\newpage

\begin{center}~\vfill\end{center}

\begin{center}\begin{tabular}{c}
\includegraphics[%
  clip,
  width=0.75\columnwidth,
  keepaspectratio]{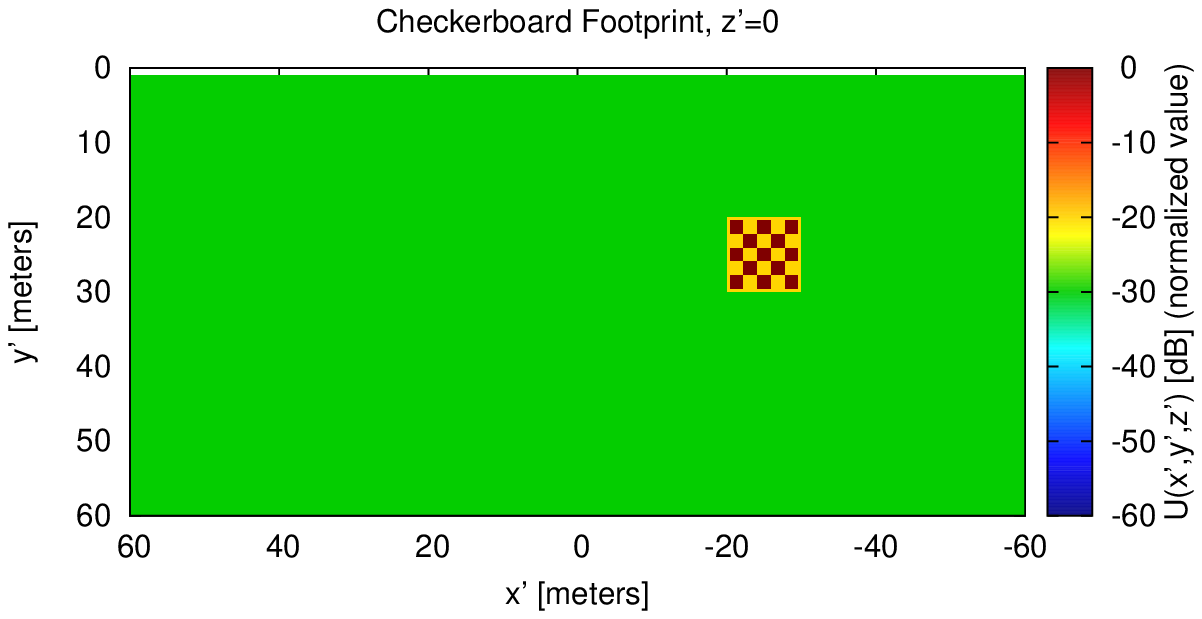}\tabularnewline
(\emph{a})\tabularnewline
\includegraphics[%
  clip,
  width=0.85\columnwidth,
  keepaspectratio]{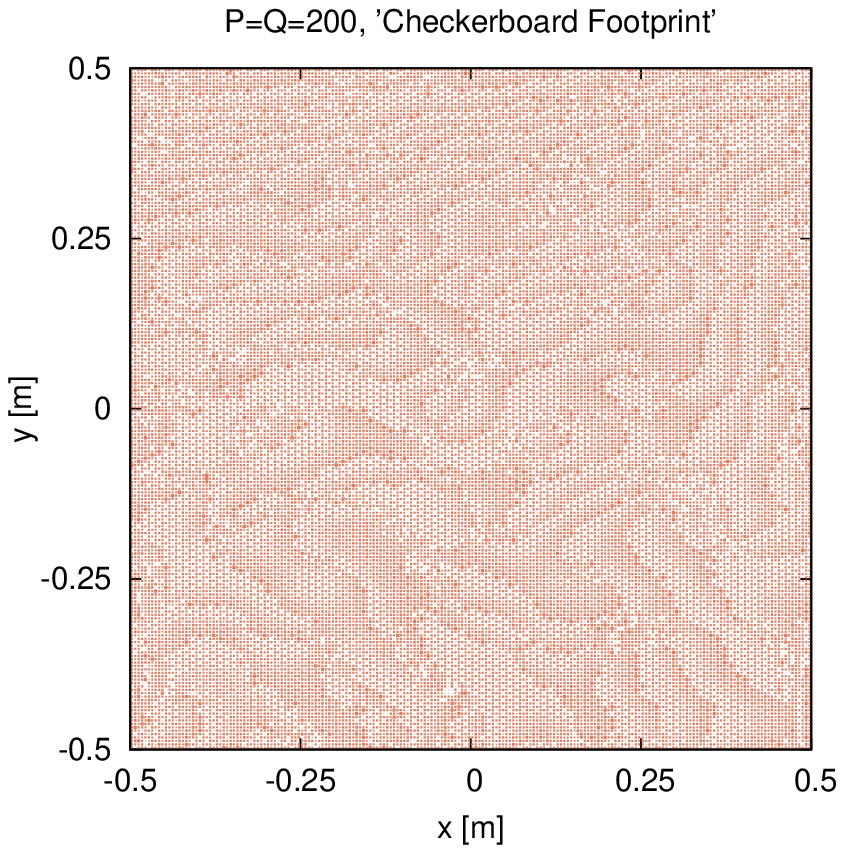}\tabularnewline
(\emph{b})\tabularnewline
\end{tabular}\end{center}

\begin{center}\vfill\end{center}

\begin{center}\textbf{Fig. 6 - G. Oliveri et} \textbf{\emph{al.}}\textbf{,}
{}``Holographic Smart \emph{EM} Skins ...''\end{center}
\newpage

\begin{center}\begin{tabular}{c}
\includegraphics[%
  clip,
  width=0.95\columnwidth,
  keepaspectratio]{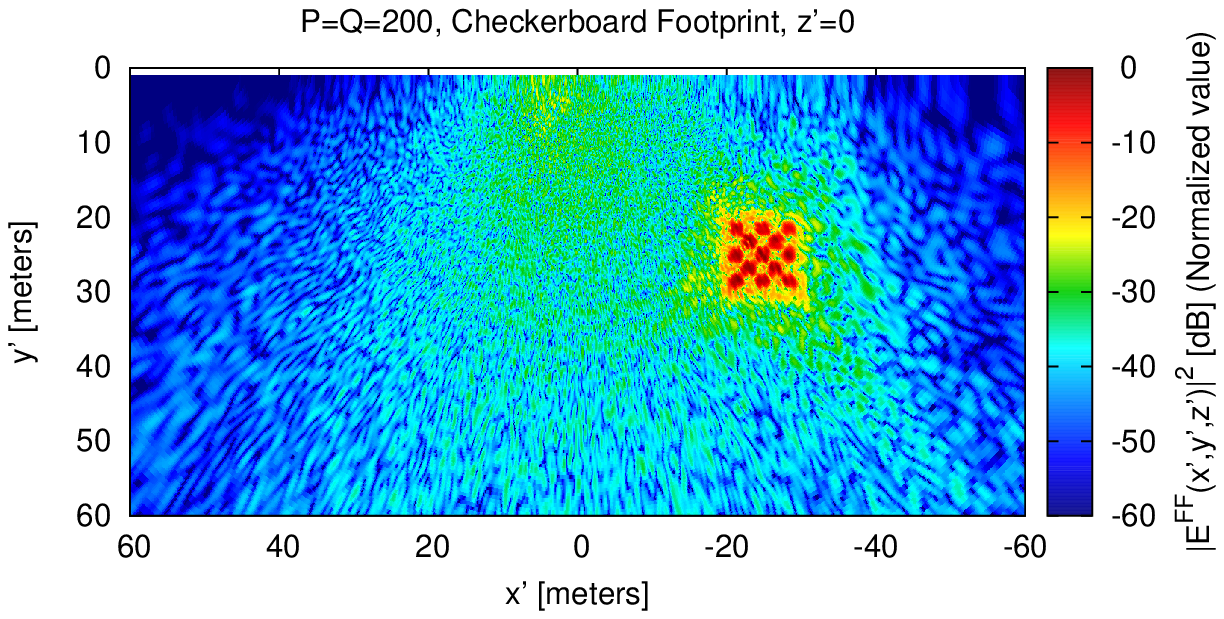}\tabularnewline
(\emph{a})\tabularnewline
\tabularnewline
\includegraphics[%
  clip,
  width=0.90\columnwidth,
  keepaspectratio]{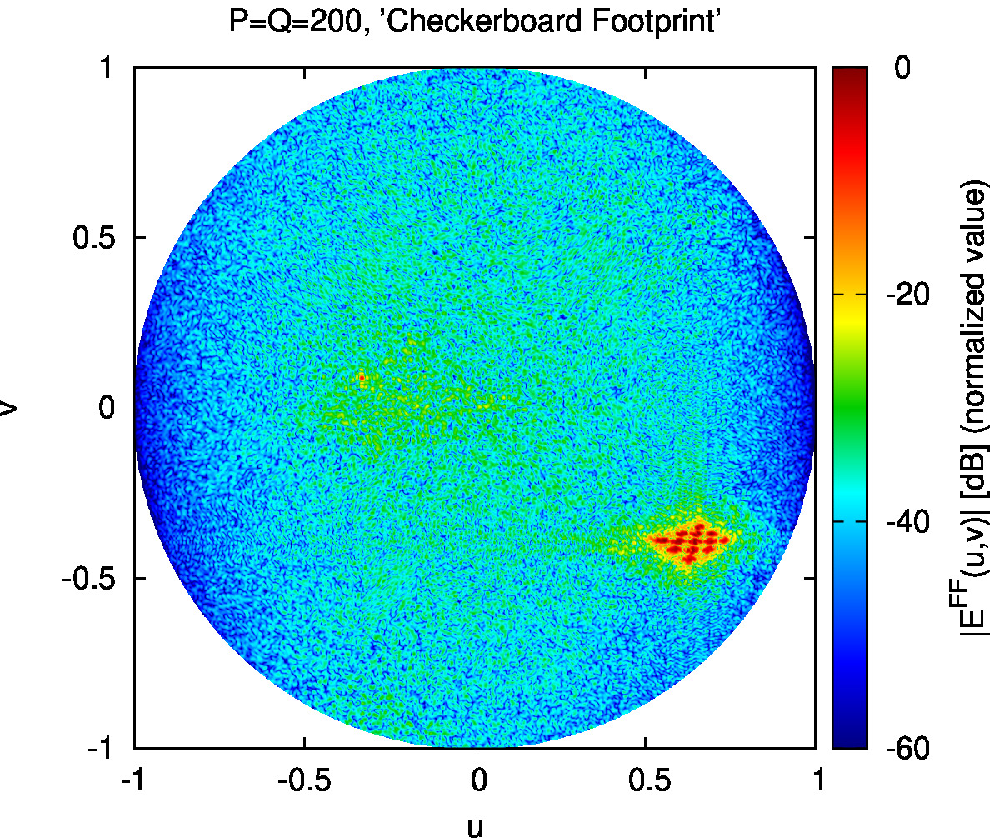}\tabularnewline
(\emph{b})\tabularnewline
\end{tabular}\end{center}

\begin{center}\vfill\end{center}

\begin{center}\textbf{Fig. 7 - G. Oliveri et} \textbf{\emph{al.}}\textbf{,}
{}``Holographic Smart \emph{EM} Skins ...''\end{center}
\newpage

\begin{center}\begin{tabular}{cc}
\multicolumn{2}{c}{\includegraphics[%
  clip,
  width=0.45\columnwidth,
  keepaspectratio]{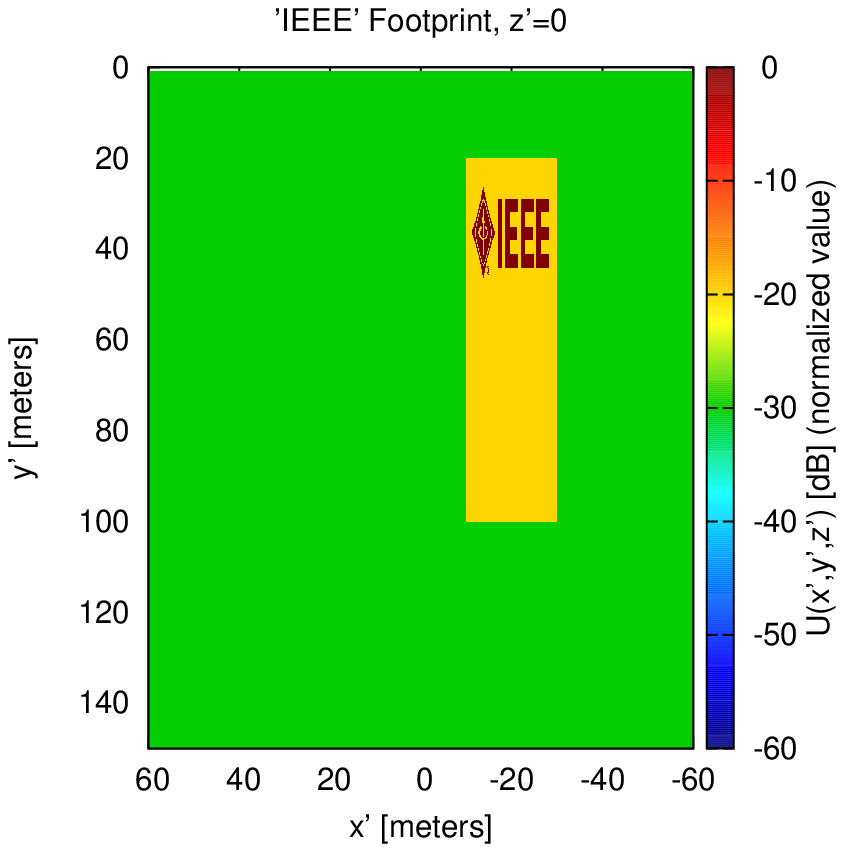}}\tabularnewline
\multicolumn{2}{c}{(\emph{a})}\tabularnewline
\multicolumn{2}{c}{\includegraphics[%
  clip,
  width=0.70\columnwidth,
  keepaspectratio]{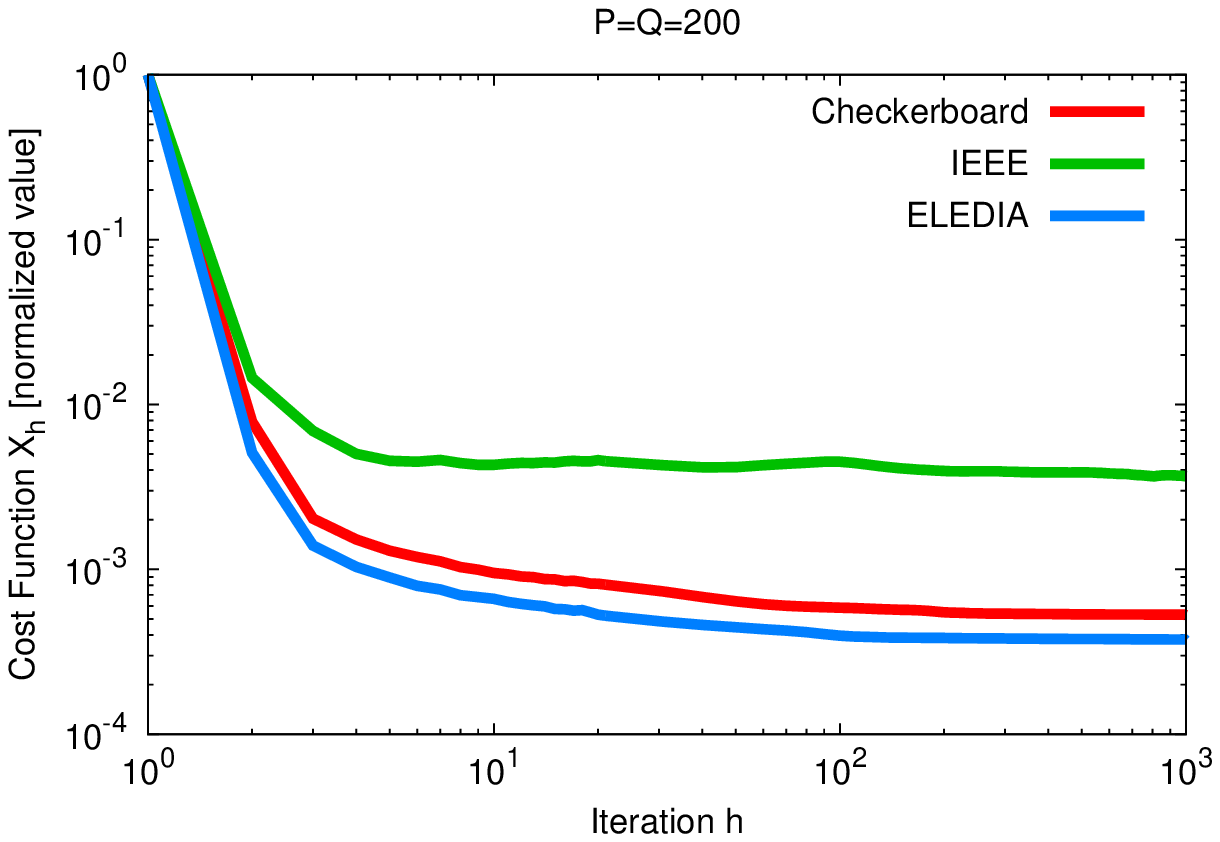}}\tabularnewline
\multicolumn{2}{c}{(\emph{b})}\tabularnewline
\multicolumn{2}{c}{\includegraphics[%
  clip,
  width=0.70\columnwidth,
  keepaspectratio]{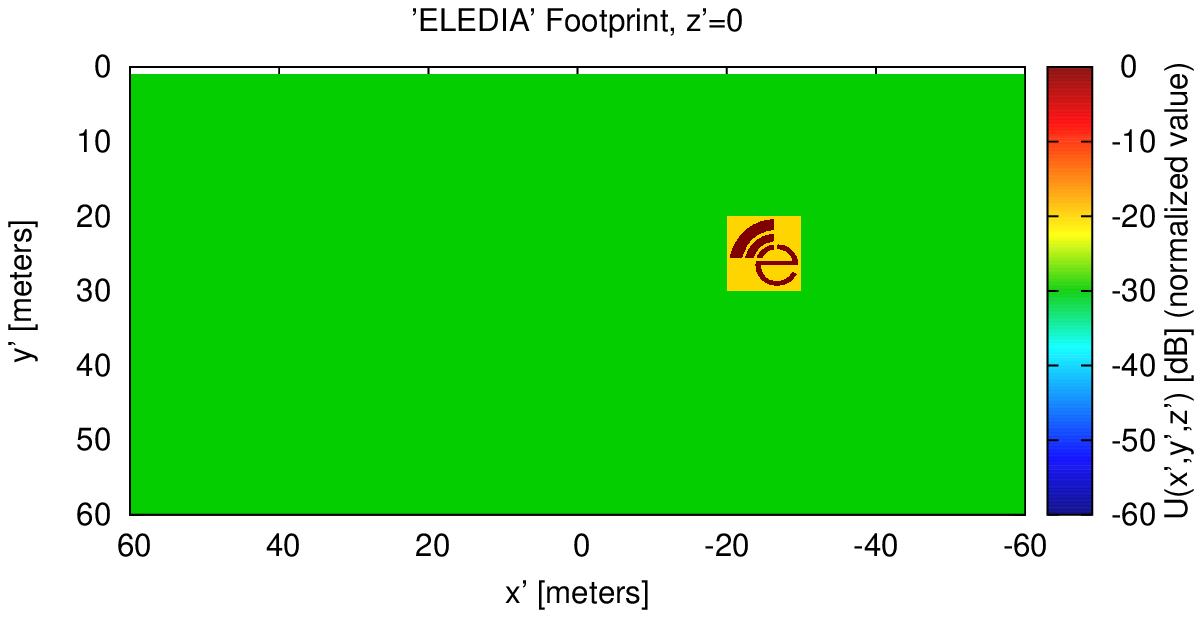}}\tabularnewline
\multicolumn{2}{c}{(\emph{c})}\tabularnewline
\end{tabular}\end{center}

\begin{center}\textbf{Fig. 8 - G. Oliveri et} \textbf{\emph{al.}}\textbf{,}
{}``Holographic Smart \emph{EM} Skins ...''\end{center}
\newpage

\begin{center}\begin{tabular}{c}
\includegraphics[%
  clip,
  width=0.65\columnwidth,
  keepaspectratio]{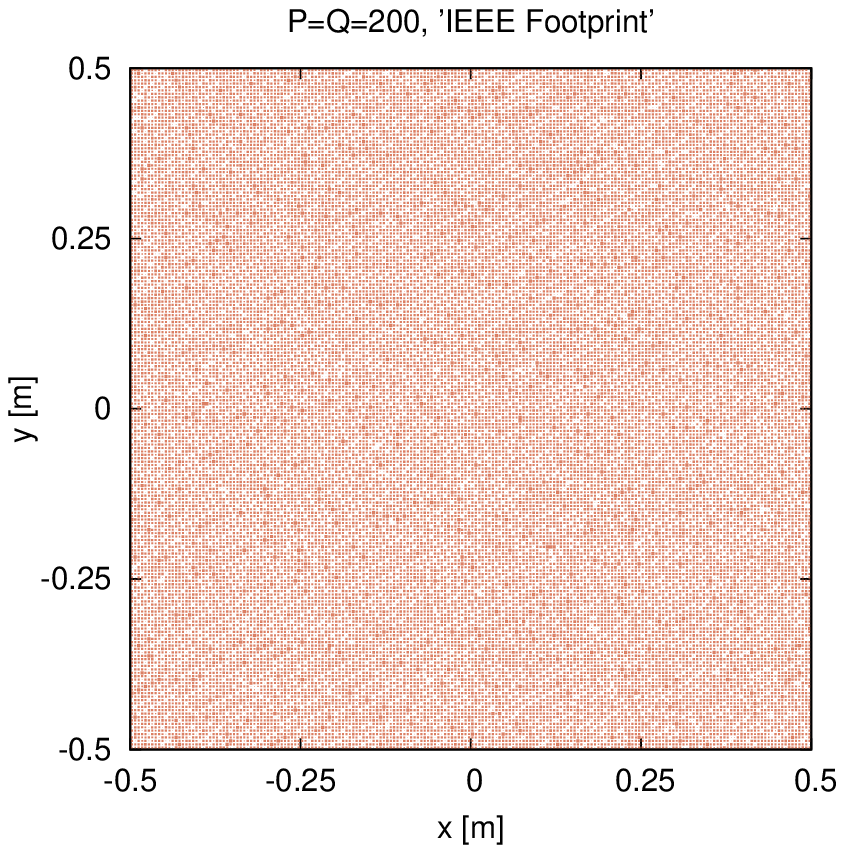}\tabularnewline
(\emph{a})\tabularnewline
\includegraphics[%
  clip,
  width=0.70\columnwidth,
  keepaspectratio]{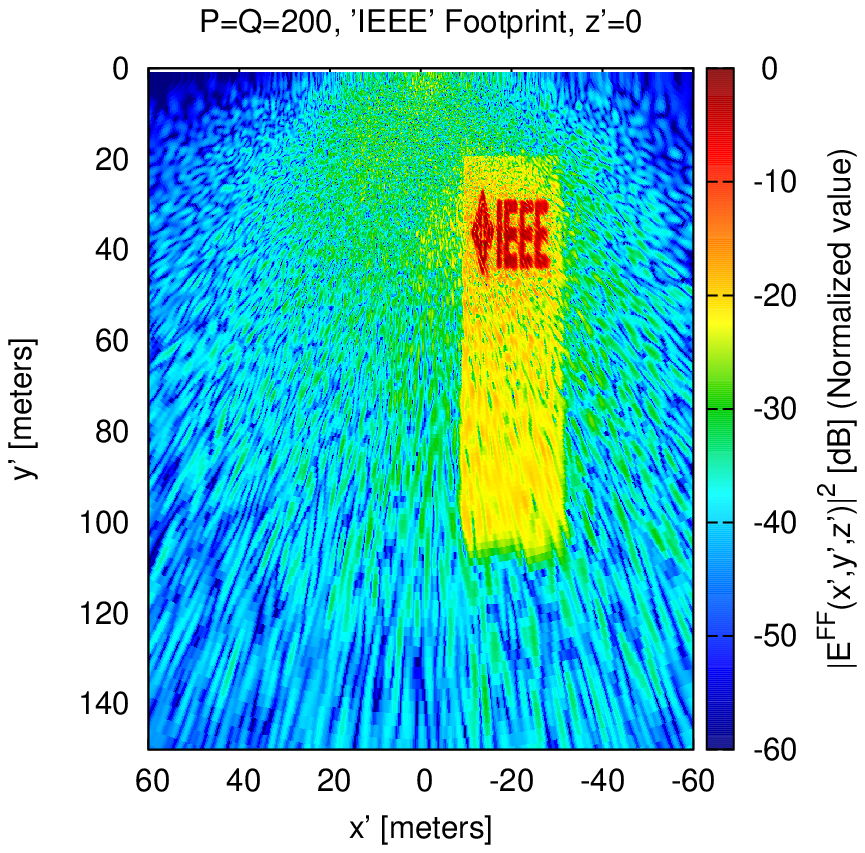}\tabularnewline
(\emph{b})\tabularnewline
\end{tabular}\end{center}

\begin{center}\textbf{Fig. 9 - G. Oliveri et} \textbf{\emph{al.}}\textbf{,}
{}``Holographic Smart \emph{EM} Skins ...''\end{center}
\newpage

\begin{center}\begin{tabular}{cc}
\includegraphics[%
  clip,
  width=0.42\columnwidth,
  keepaspectratio]{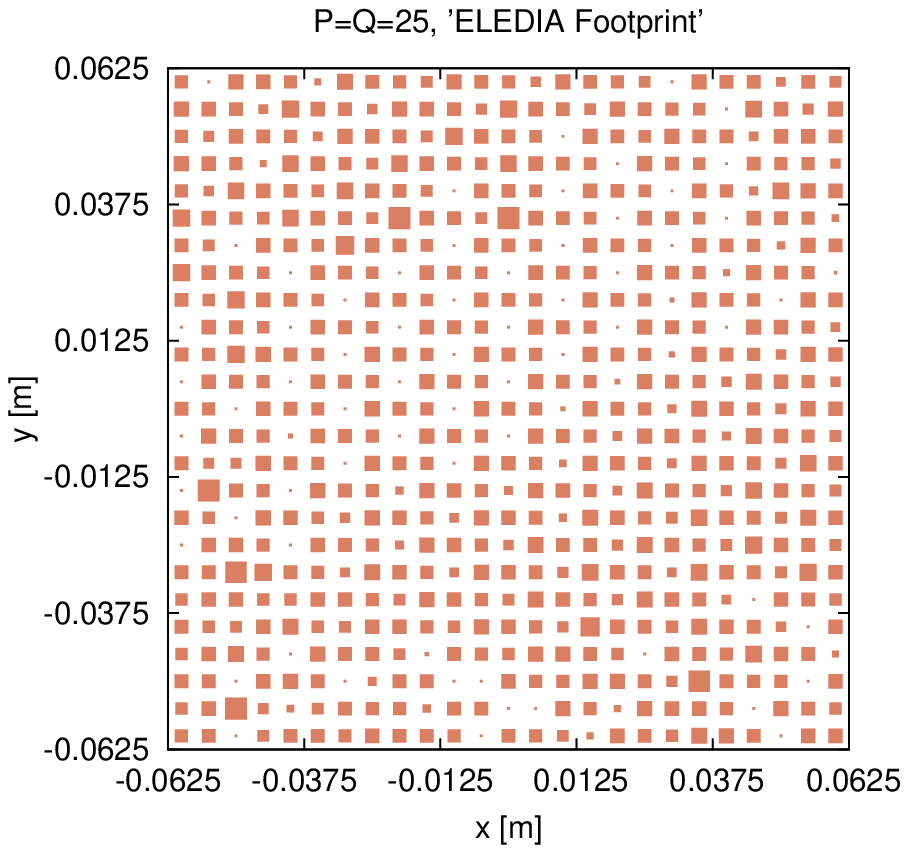}&
\includegraphics[%
  clip,
  width=0.42\columnwidth,
  keepaspectratio]{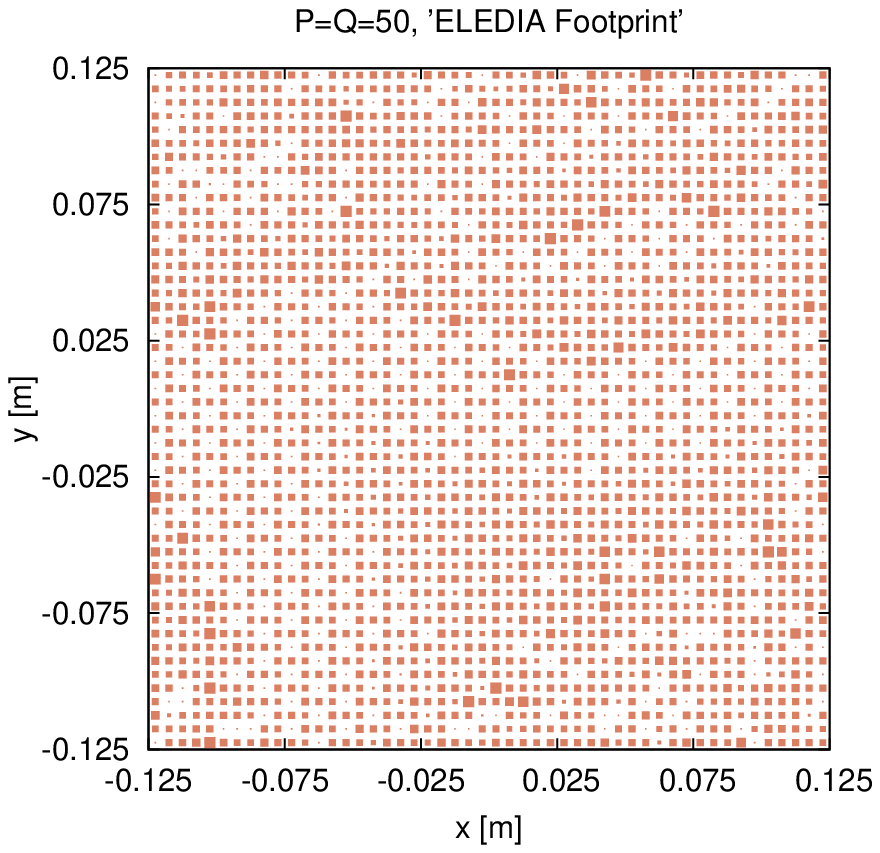}\tabularnewline
(\emph{a})&
(\emph{b})\tabularnewline
\includegraphics[%
  clip,
  width=0.42\columnwidth,
  keepaspectratio]{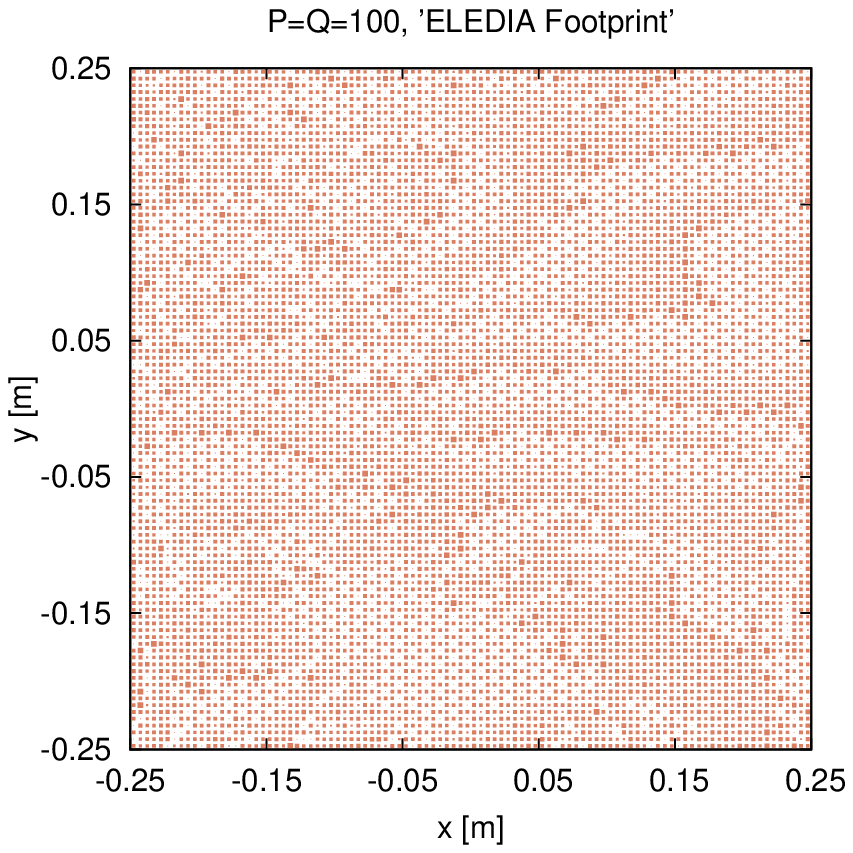}&
\includegraphics[%
  clip,
  width=0.42\columnwidth,
  keepaspectratio]{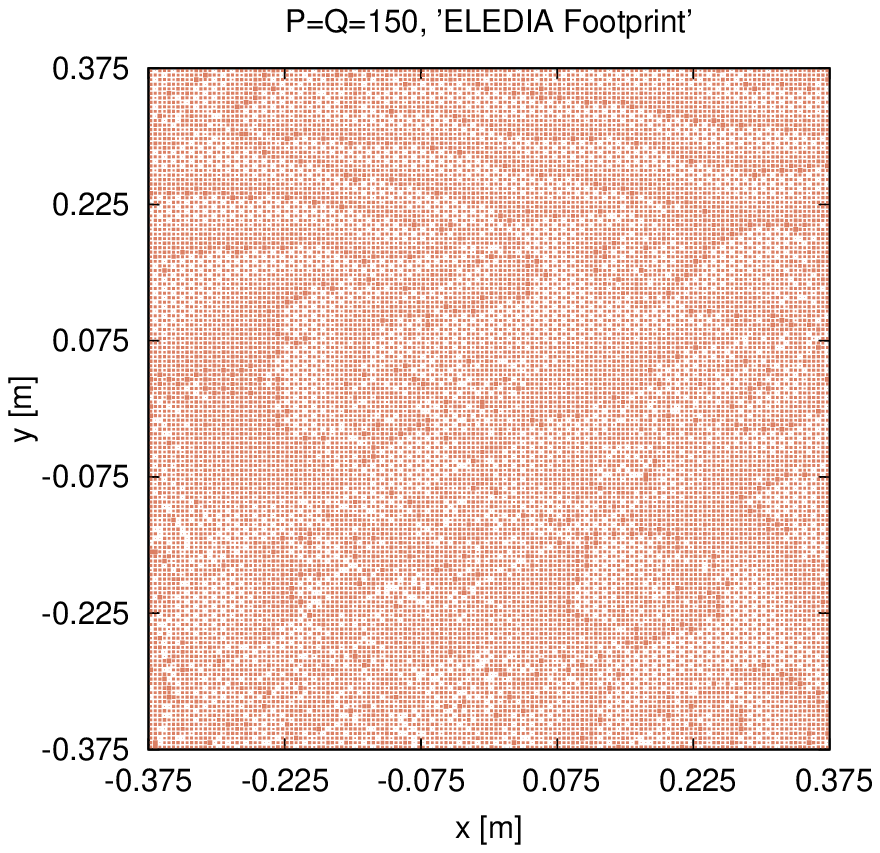}\tabularnewline
(\emph{c})&
(\emph{d})\tabularnewline
\includegraphics[%
  clip,
  width=0.42\columnwidth,
  keepaspectratio]{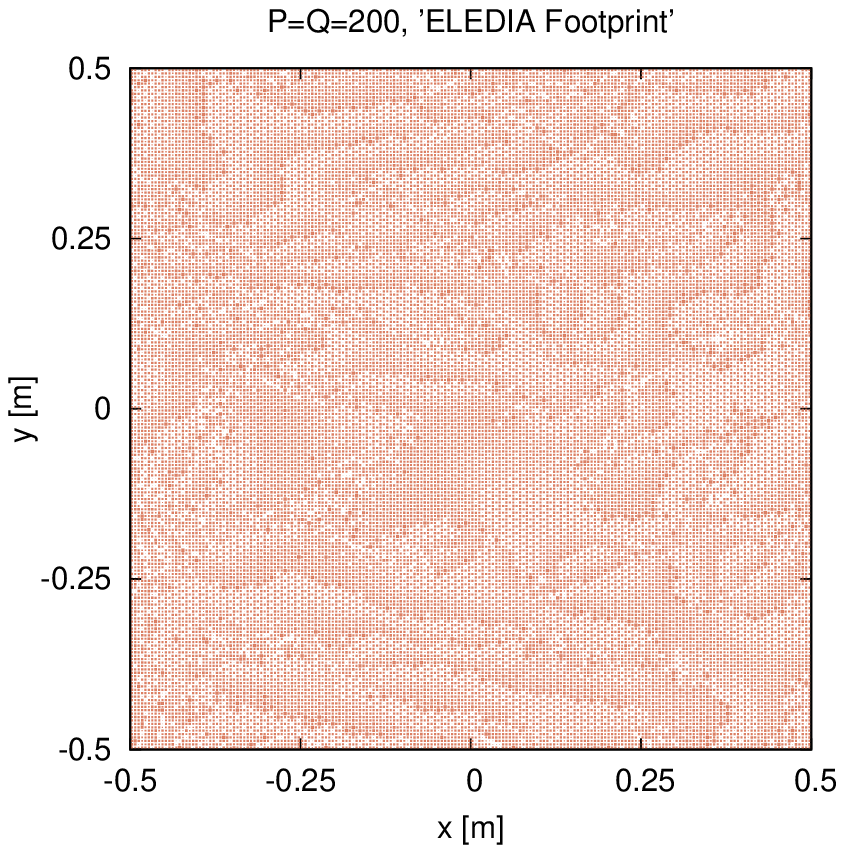}&
\includegraphics[%
  clip,
  width=0.42\columnwidth,
  keepaspectratio]{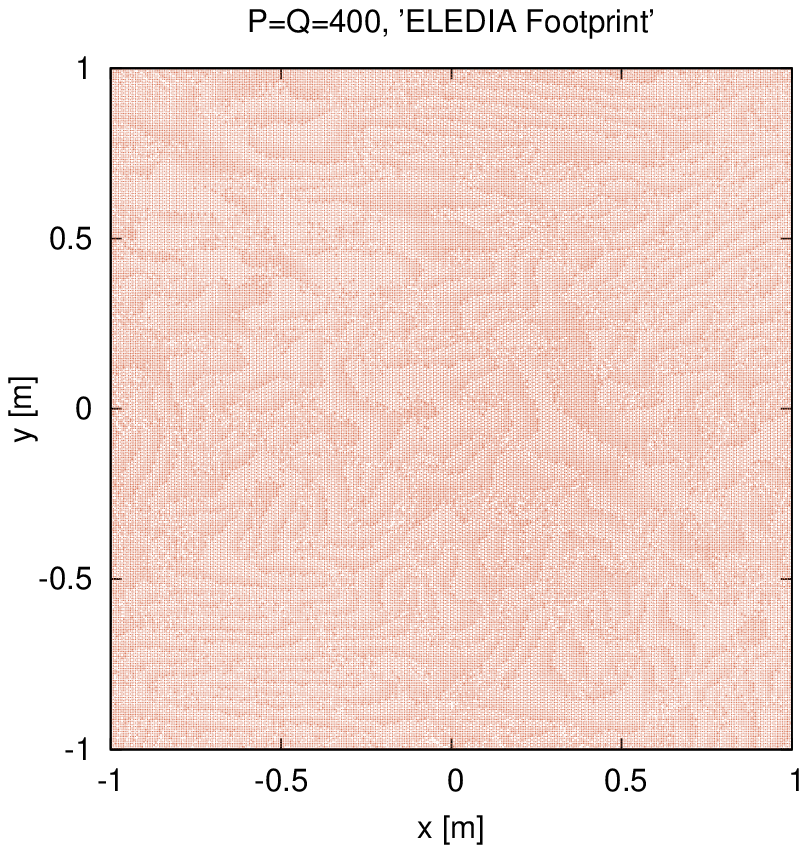}\tabularnewline
(\emph{e})&
(\emph{f})\tabularnewline
\end{tabular}\end{center}

\begin{center}\textbf{Fig. 10 - G. Oliveri et} \textbf{\emph{al.}}\textbf{,}
{}``Holographic Smart \emph{EM} Skins ...''\end{center}
\newpage

\begin{center}~\vfill\end{center}

\begin{center}\begin{tabular}{cc}
\includegraphics[%
  clip,
  width=0.48\columnwidth,
  keepaspectratio]{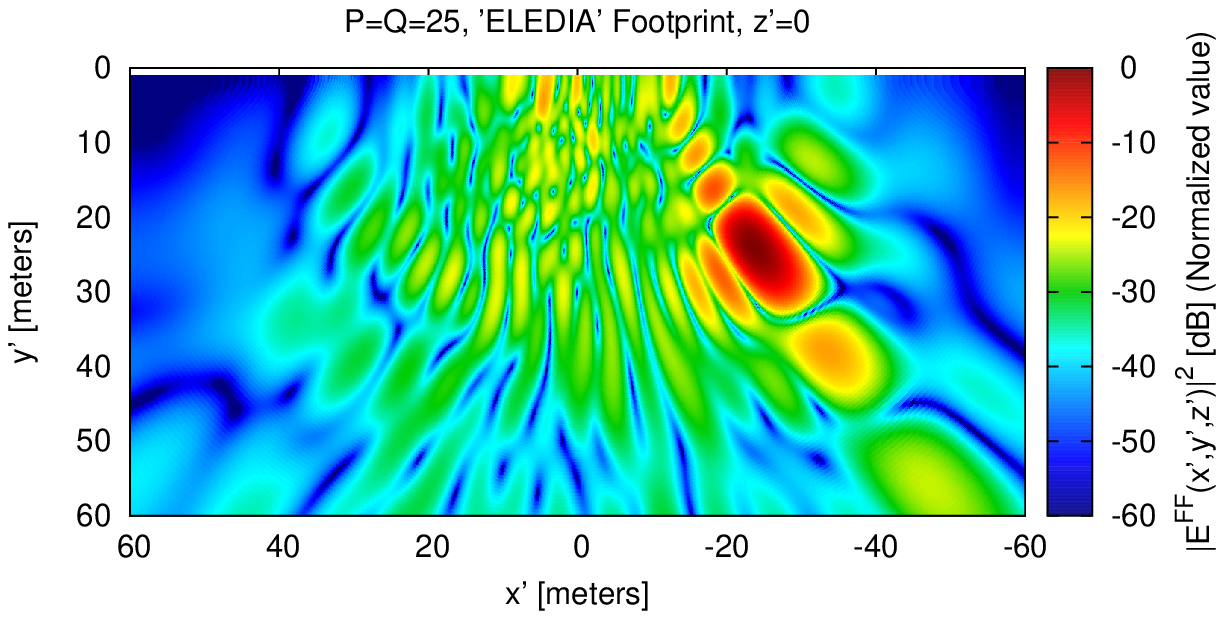}&
\includegraphics[%
  clip,
  width=0.48\columnwidth,
  keepaspectratio]{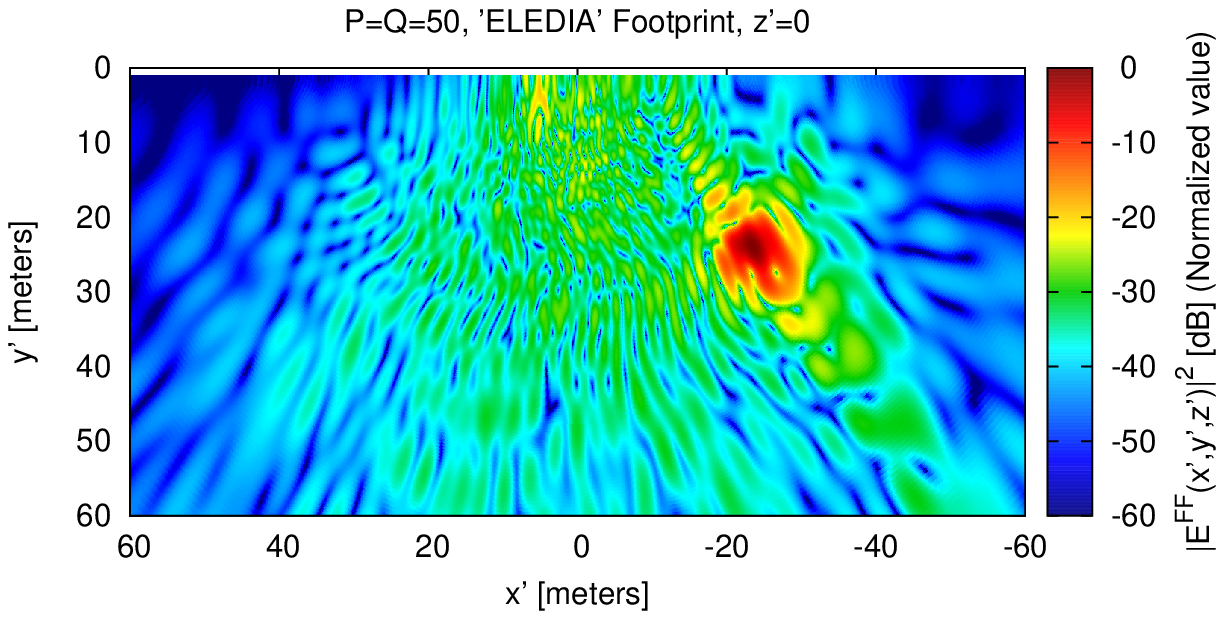}\tabularnewline
(\emph{a})&
(\emph{b})\tabularnewline
\includegraphics[%
  clip,
  width=0.48\columnwidth,
  keepaspectratio]{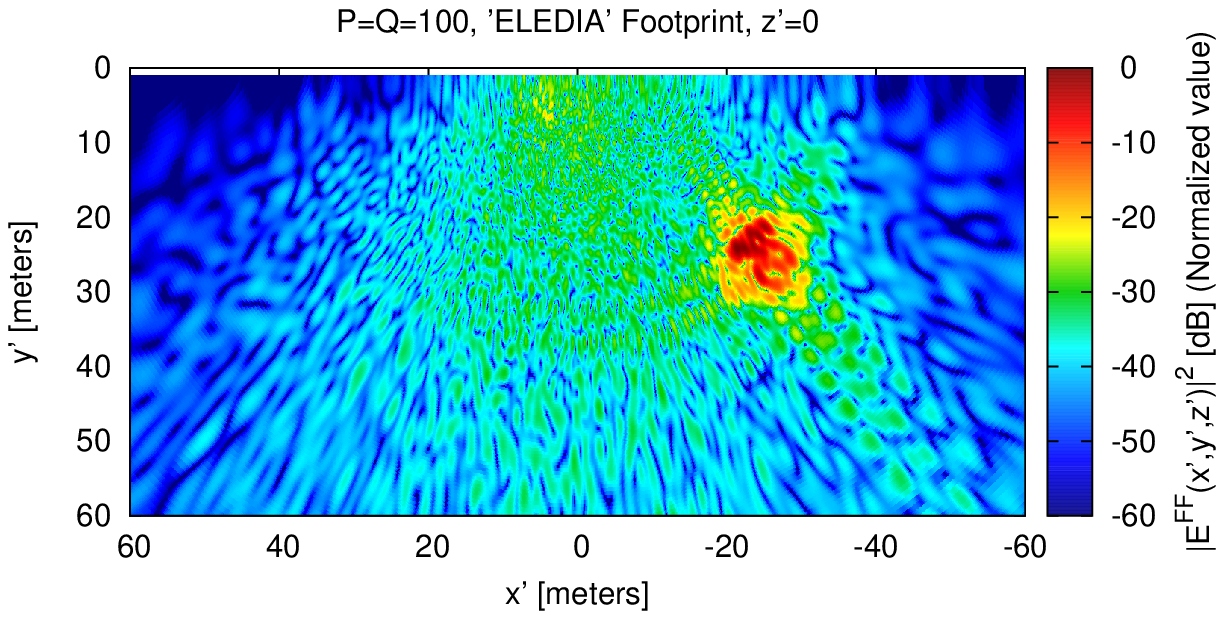}&
\includegraphics[%
  clip,
  width=0.48\columnwidth,
  keepaspectratio]{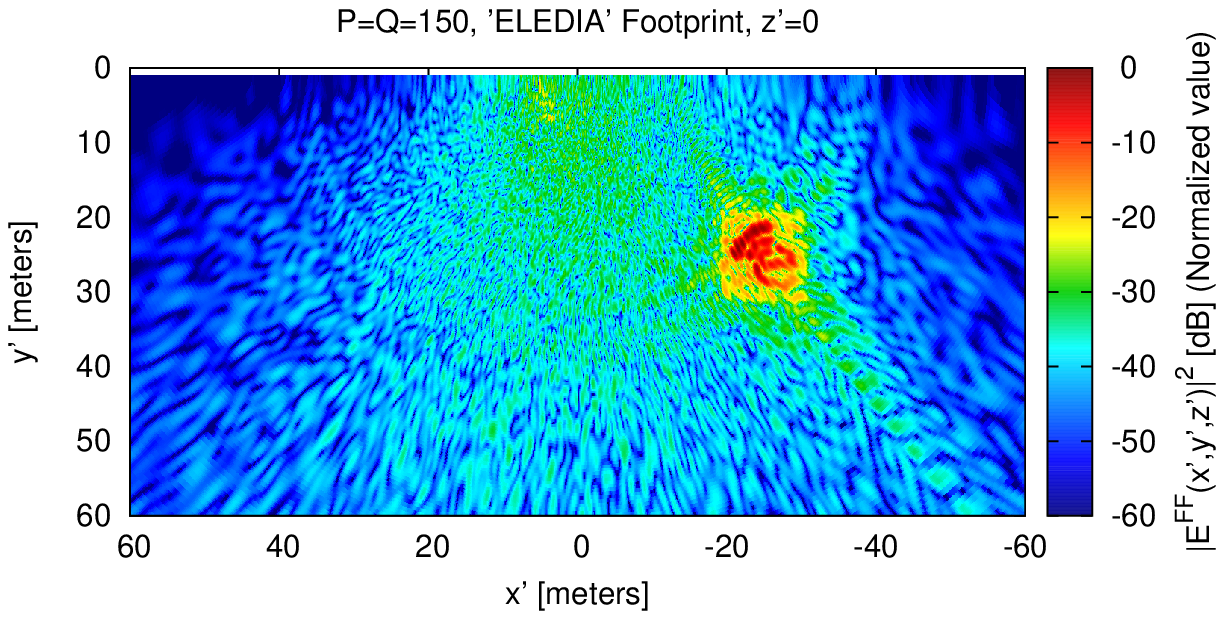}\tabularnewline
(\emph{c})&
(\emph{d})\tabularnewline
\includegraphics[%
  clip,
  width=0.48\columnwidth,
  keepaspectratio]{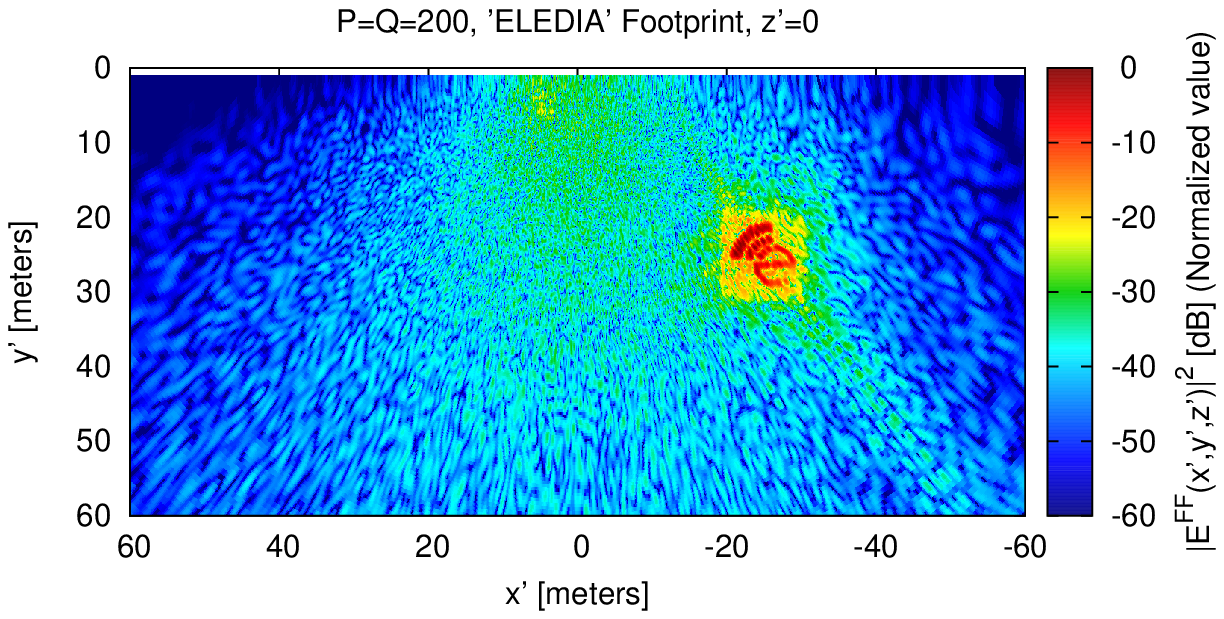}&
\includegraphics[%
  clip,
  width=0.48\columnwidth,
  keepaspectratio]{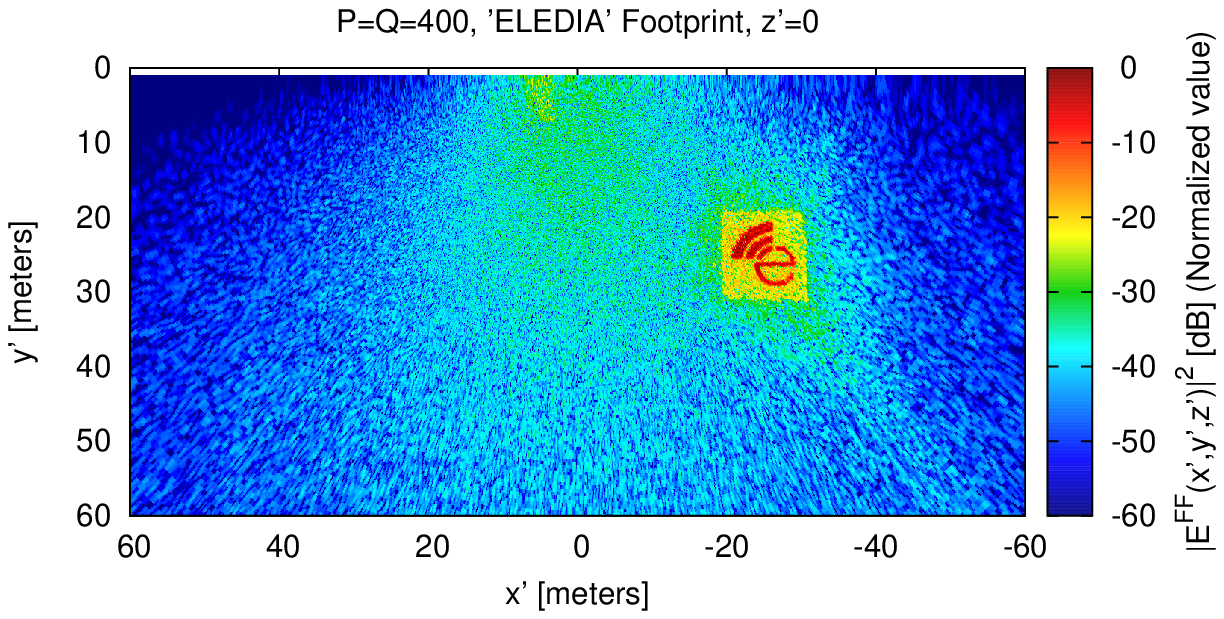}\tabularnewline
(\emph{e})&
(\emph{f})\tabularnewline
\end{tabular}\end{center}

\begin{center}\vfill\end{center}

\begin{center}\textbf{Fig. 11 - G. Oliveri et} \textbf{\emph{al.}}\textbf{,}
{}``Holographic Smart \emph{EM} Skins ...''\end{center}
\newpage

\begin{center}\begin{tabular}{c}
\includegraphics[%
  clip,
  width=0.95\columnwidth,
  keepaspectratio]{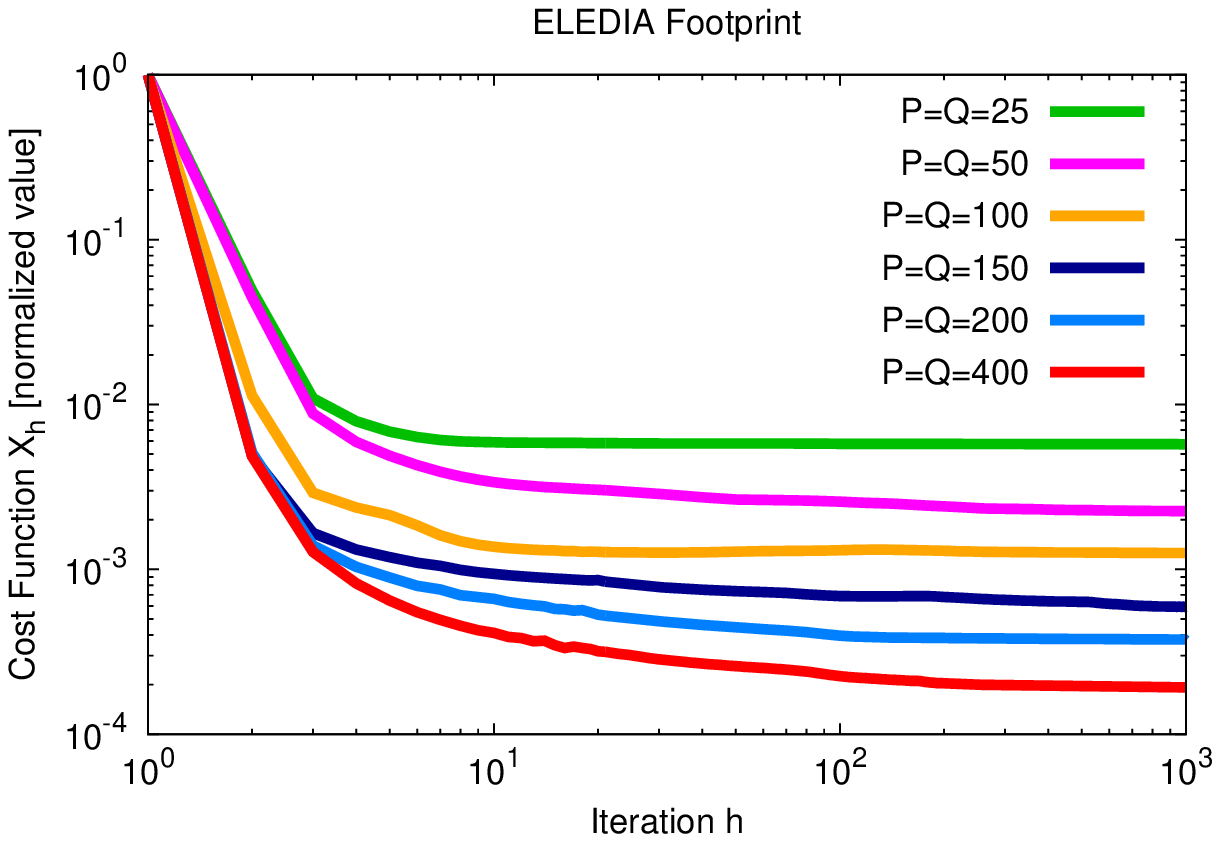}\tabularnewline
(\emph{a})\tabularnewline
\includegraphics[%
  clip,
  width=0.95\columnwidth,
  keepaspectratio]{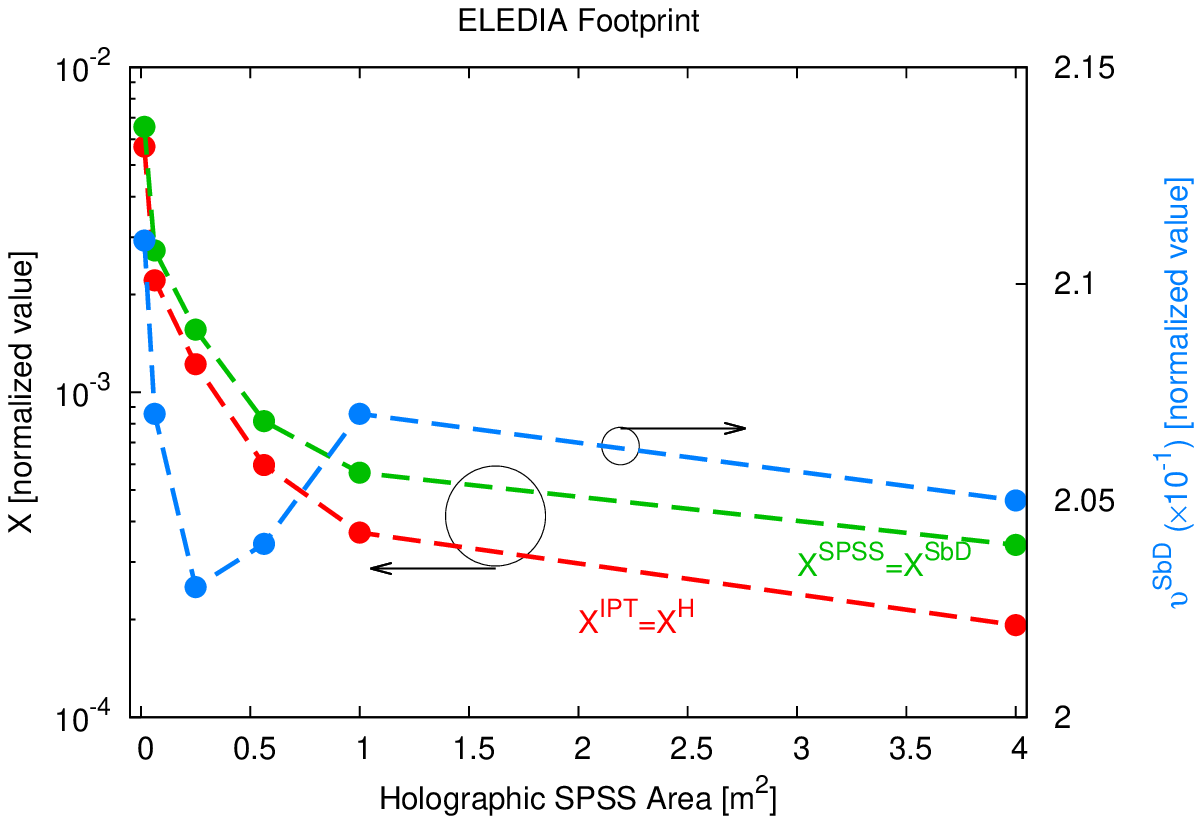}\tabularnewline
(\emph{b})\tabularnewline
\end{tabular}\end{center}

\begin{center}\textbf{Fig. 12 - G. Oliveri et} \textbf{\emph{al.}}\textbf{,}
{}``Holographic Smart \emph{EM} Skins ...''\end{center}
\newpage

\begin{center}~\vfill\end{center}

\begin{center}\begin{sideways}
\begin{tabular}{|c|c||c|c|c||c|c|c|}
\hline 
\emph{Footprint Name}&
\emph{Footprint Mask}&
$P\times Q$&
$\Delta t^{IPT}$ {[}s{]}&
$\Delta t^{SbD}$ {[}s{]}&
$\mathcal{X}^{IPT}$&
$\upsilon^{SbD}$&
$\mathcal{X}^{SPSS}$\tabularnewline
\hline
\hline 
\emph{Square}&
Fig. 3(\emph{a})&
$200\times200$&
$2.31\times10^{2}$&
$9.30$&
$6.44\times10^{-4}$&
$2.05\times10^{-1}$&
$1.08\times10^{-3}$\tabularnewline
\hline 
\emph{Checkerboard}&
Fig. 6(\emph{a})&
$200\times200$&
$2.37\times10^{2}$&
$9.08$&
$5.27\times10^{-4}$&
$2.04\times10^{-1}$&
$7.78\times10^{-4}$\tabularnewline
\hline 
\emph{IEEE}&
Fig. 8(\emph{b})&
$200\times200$&
$1.89\times10^{2}$&
$8.82$&
$3.65\times10^{-3}$&
$2.06\times10^{-1}$&
$4.84\times10^{-3}$\tabularnewline
\hline 
\emph{ELEDIA}&
Fig. 8(\emph{c})&
$25\times25$&
$1.22\times10^{2}$&
$1.64\times10^{-1}$&
$5.70\times10^{-3}$&
$2.11\times10^{-1}$&
$6.55\times10^{-3}$\tabularnewline
\hline 
\emph{ELEDIA}&
Fig. 8(\emph{c})&
$50\times50$&
$1.46\times10^{2}$&
$6.20\times10^{-1}$&
$2.21\times10^{-3}$&
$2.07\times10^{-1}$&
$2.72\times10^{-3}$\tabularnewline
\hline 
\emph{ELEDIA}&
Fig. 8(\emph{c})&
$100\times100$&
$1.53\times10^{2}$&
$2.45$&
$1.22\times10^{-3}$&
$2.03\times10^{-1}$&
$1.55\times10^{-3}$\tabularnewline
\hline 
\emph{ELEDIA}&
Fig. 8(\emph{c})&
$150\times150$&
$1.64\times10^{2}$&
$6.35$&
$5.97\times10^{-4}$&
$2.04\times10^{-1}$&
$8.15\times10^{-4}$\tabularnewline
\hline 
\emph{ELEDIA}&
Fig. 8(\emph{c})&
$200\times200$&
$2.66\times10^{2}$&
$9.31$&
$3.70\times10^{-4}$&
$2.07\times10^{-1}$&
$5.65\times10^{-4}$\tabularnewline
\hline 
\emph{ELEDIA}&
Fig. 8(\emph{c})&
$400\times400$&
$1.03\times10^{3}$&
$3.61\times10^{1}$&
$1.92\times10^{-4}$&
$2.05\times10^{-1}$&
$3.38\times10^{-4}$\tabularnewline
\hline
\end{tabular}
\end{sideways}\end{center}

\begin{center}\vfill\end{center}

\begin{center}\textbf{Table I - G. Oliveri et} \textbf{\emph{al.}}\textbf{,}
{}``Holographic Smart \emph{EM} Skins ...''\end{center}
\end{document}